\documentclass[12pt]{article}
\usepackage[utf8]{inputenc}
\usepackage{graphicx}
\usepackage{color}
\usepackage{supertabular}
\usepackage{ltxtable}
\usepackage{authblk}

\usepackage[superscript]{cite}

\RequirePackage{cite}
\RequirePackage{times}
\RequirePackage{fullpage}
\RequirePackage{ifthen}
\usepackage{xpatch}

\def\ferre{{\tt FER\reflectbox{R}E}}

\def\kmso{\hbox{${\rm km}\:{\rm s}^{-1}$}}
\def\kms{\hbox{${\rm km}\:{\rm s}^{-1}\,$}}

\newcommand{\aap}{Astron. Astrophys.}   
\newcommand{\aaps}{Astron. Astrophys. Suppl.}   
\newcommand{\pasp}{Publ. Astron. Soc. Pac.}   
\newcommand{\apj}{Astrophys. J.}   
\newcommand{\apjl}{Astrophys. J. Lett.}   
\newcommand{\araa}{Annu. Rev. Astron. Astrophys.}   
\newcommand{\aj}{Astron. J.}   
\newcommand{\mnras}{Mon. Not. R. Astron. Soc.}   
\newcommand{\icarus}{Icarus}   
\newcommand{\nat}{Nature} 
\newcommand{\grl}{Geophys. Res. Lett.}

\newcommand{\spacing}[1]{\renewcommand{\baselinestretch}{#1}\large\normalsize}
\spacing{1.5}

\def\@maketitle{%
  \newpage\spacing{1}\setlength{\parskip}{12pt}%
    {\Large\bfseries\noindent\sloppy \textsf{\@title} \par}%
    {\noindent\sloppy \@author}%
}

\title{Rapid contraction of giant planets orbiting the 20 million-years old star V1298 Tau}

\author{A. Su\'{a}rez Mascare\~{n}o$^{1,2}$, 
M. Damasso$^3$,
N. Lodieu$^{1,2}$,
A. Sozzetti$^{3}$,
V. J. S. Béjar$^{1,2}$, 
S. Benatti$^{4}$,
M. R. Zapatero Osorio$^{5}$,
G. Micela$^{4}$,
R. Rebolo$^{1,2,6}$,
S. Desidera$^{7}$,
F. Murgas$^{1,2}$,
R. Claudi$^{7}$, 
J. I. González Hernández$^{1,2}$,
L. Malavolta$^{7,8}$,
C. del Burgo$^{9, 1}$,
V. D'Orazi$^{7, 10}$,
P. J. Amado$^{11}$,
D. Locci$^{4}$,
H. M. Tabernero$^{5, 12}$,
F. Marzari$^{8}$, 
D. S. Aguado $^{13}$,
D. Turrini$^{14}$,
C. Cardona Guillén$^{1,2}$,
B. Toledo-Padrón$^{1,2}$,
A. Maggio $^{4}$,
J. Aceituno$^{15, 11}$,
F. F. Bauer$^{11}$, 
J. A. Caballero$^{16}$,
P. Chinchilla$^{17,1}$,
E. Esparza-Borges $^{2}$,
E. González-Álvarez$^{5}$,
T. Granzer$^{18}$,
R. Luque$^{11}$,
E. L. Martín$^{1,2, 6}$,
G. Nowak$^{1,2}$,
M. Oshagh$^{1,2}$,
E. Pallé$^{1,2}$,
H. Parviainen$^{1,2}$,
A. Quirrenbach$^{19}$,
A. Reiners $^{20}$,
I. Ribas $^{21, 22}$, 
K. G. Strassmeier$^{18}$,
M. Weber$^{18}$ 
and M. Mallonn$^{18}$}
\begin{document}

\maketitle

\newenvironment{affiliations}{%
    \setcounter{enumi}{1}%
    \setlength{\parindent}{0in}%
    \slshape\sloppy%
    \begin{list}{\upshape$^{\arabic{enumi}}$}{%
        \usecounter{enumi}%
        \setlength{\leftmargin}{0in}%
        \setlength{\topsep}{0in}%
        \setlength{\labelsep}{0in}%
        \setlength{\labelwidth}{0in}%
        \setlength{\listparindent}{0in}%
        \setlength{\itemsep}{0ex}%
        \setlength{\parsep}{0in}%
        }
    }{\end{list}\par\vspace{12pt}}

\renewenvironment{abstract}{%
    \setlength{\parindent}{0in}%
    \setlength{\parskip}{0in}%
    \bfseries%
    }{\par\vspace{-6pt}}

\newenvironment{addendum}{%
    \setlength{\parindent}{0in}%
    \small%
    \begin{list}{Acknowledgements}{%
        \setlength{\leftmargin}{0in}%
        \setlength{\listparindent}{0in}%
        \setlength{\labelsep}{0em}%
        \setlength{\labelwidth}{0in}%
        \setlength{\itemsep}{12pt}%
        \let\makelabel\addendumlabel}
    }
    {\end{list}\normalsize}

\newcommand*{\addendumlabel}[1]{\textbf{#1}\hspace{1em}}

\begin{affiliations}
\item Instituto de Astrof\'{i}sica de Canarias, E-38205 La Laguna, Tenerife, Spain. 
\item Departamento de Astrof\`{i}sica, Universidad de La Laguna, E-38206 La Laguna, Tenerife, Spain.
\item Istituto Nazionale di Astrofisica -- Osservatorio Astrofisico di Torino, I-10025 Pino Torinese (TO), Italy .
\item Istituto Nazionale di Astrofisica -- Osservatorio Astronomico di Palermo, I-90134 Palermo, Italy.
\item Centro de Astrobiología (Consejo Superior de Investigaciones Científicas – Instituto Nacional de Técnica Aeroespacial), E-28850 Torrejón de Ardoz Madrid, Spain. 
\item Consejo Superior de Investigaciones Científicas, E-28006 Madrid, Spain.
 \item Istituto Nazionale di Astrofisica – Osservatorio Astronomico di Padova, I-35122 Padova, Italy.
\item Dipartimento di Fisica e Astronomia "Galileo Galilei", Università degli Studi di Padova, I-35122, Padova, Italy.
\item Instituto Nacional de Astrofísica, Óptica y Electrónica, 72840 Puebla, Mexico.
\item School of Physics and Astronomy, Monash University, VIC-3800 Melbourne, Australia.
\item Consejo Superior de Investigaciones Científicas -- Instituto de Astrofísica de Andalucía, E-18008 Granada, Spain.
\item Instituto de Astrofísica e Ciências do Espaço, Universidade do Porto, CAUP, P-4150-762 Porto, Portugal.
\item Institute of Astronomy, University of Cambridge, Cambridge CB3 0HA, UK.
\item Institute for Space Astrophysics and Planetology (Istituto Nazionale di Astrofisica, Associazione Italiana Studenti di Fisica), I-00133 Rome, Italy.
\item Centro Astronómico Hispano Alemán, E-04550 Gérgal, Almería, Spain.
\item Centro de Astrobiología (Consejo Superior de Investigaciones Científicas – Instituto Nacional de Técnica Aeroespacial), ESAC, E-28692 Villanueva de la Cañada, Madrid, Spain.
\item Astrobiology Research Unit, Universit\'e de Li\`ege, 19C All\`ee du 6 Ao\^ut, 4000 Li\`ege, Belgium
\item Leibniz-Institute for Astrophysics Potsdam, D-14482 Potsdam, Germany .
\item Landessternwarte, Zentrum für Astronomie der Universität Heidelberg, D-69117 Heidelberg, Germany.
\item Institut für Astrophysik, Georg-August-Universität Göttingen, D-37077 Göttingen, Germany.
\item Institut de Ciències de l’Espai (Instituto de Ciencias del Espacio, Consejo Superior de Investigaciones Científicas), Campus UAB, E-08193 Bellaterra, Spain.
\item Institut d’Estudis Espacials de Catalunya, E-08034 Barcelona, Spain.

\end{affiliations}

\begin{abstract}

Current theories of planetary evolution predict that infant giant planets have large radii and very low densities before they slowly contract to reach their final size after about several hundred million years\cite{Mordasini2012,DAngelo2021}. These theoretical expectations remain untested to date, despite the increasing number of exoplanetary discoveries, as the detection and characterisation of very young planets is extremely challenging due to the intense stellar activity of their host stars\cite{Donati2016, Damasso2020}. However, the recent discoveries of young planetary transiting systems allow to place initial constraints on evolutionary models\cite{David16,Plavchan2020, Klein2021}. With an estimated age of 20 million years, V1298\,Tau is one of the youngest solar-type stars known to host transiting planets: it harbours a multiple system composed of two Neptune-sized, one Saturn-sized, and one Jupiter-sized planets\cite{David2019, DavidT2019}.  Here we report the analysis of an intense radial velocity campaign, revealing the presence of two periodic signals compatible with the orbits of two of its planets. We find that planet b, with an orbital period of 24 days, has a mass of 0.64 Jupiter masses and a density similar to the giant planets of the Solar System and other known giant exoplanets with significantly older ages\cite{Brahm2016, Mancini2016}. Planet  e, with an orbital period of 40 days, has a mass of 1.16 Jupiter masses and a density larger than most giant exoplanets. This is unexpected for planets at such a young age and suggests that some giant planets might evolve and contract faster than anticipated, thus challenging current models of planetary evolution.

\end{abstract}

 \noindent V1298\,Tau is a relatively bright ($V$ = 10.1) very young K1 star with a mass of 1.170$\pm$0.060 M$_{\odot}$, a radius of 1.278$\pm$0.070 R$_{\odot}$, an effective temperature of 5050 $\pm$ 100 K, and solar metallicity (see Table~\ref{stellar_params}). It is the physical companion of the G2 star HD 284154. Based on their position in color-magnitude diagrams, isochrone inference, rotation period, lithium abundance, and X-ray emission, the pair belongs to the Group 29 stellar association\cite{Oh2017} and has an age of 20 $\pm$ 10 Myr.

  \noindent V1298\,Tau  was observed by Kepler's ``Second Light" K2 mission\cite{Howell2014}. The analysis of the K2 data, covering 71 days of continuous observations, revealed the presence of four transiting planets in the system\cite{DavidT2019}. The three inner planets (b,  c, and  d) were determined to have orbital periods of 24.1396 $\pm$ 0.0018, 8.24958 $\pm$ 0.00072 and 12.4032 $\pm$ 0.0015 days,  and radii of 0.916$^{+0.052}_{-0.047}$, 0.499$^{+0.032}_{-0.029}$, and 0.572$^{+0.040}_{-0.035}$ R$_{\rm Jup}$ (i.e. Jupiter radii).  The fourth planet, e, was identified with only a single transit event, with a radius of 0.780$^{+0.075}_{-0.064}$ R$_{\rm Jup}$ and orbital period estimated to be between 40 and 120 days. A previous study constrained the mass of V1298\,Tau b to be smaller than 2.2 M$_{\rm Jup}$\cite{Beichman2019} (i.e. Jupiter masses). 

 \noindent To measure the planetary masses, we performed an intensive spectroscopic campaign collecting more than 260 radial velocity (RV) measurements of V1298\,Tau, using the high spectral resolution spectrographs HARPS-N, CARMENES, SES, and HERMES between April 2019 and April 2020. We derived RVs with a median internal uncertainty (1$\sigma$) of 9 m~s$^{-1}$ (HARPS-N) , 15 m~s$^{-1}$ (CARMENES), 50 m~s$^{-1}$ (HERMES), and 117 m~s$^{-1}$ (SES), partially caused by the rapid rotation of this young star. The combination of data coming from all four spectrographs proved convenient for monitoring the stellar activity, which causes a dispersion of the measurements of more than 200 m~s$^{-1}$, as expected given the young age of the star. However, the different wavelength coverage of each instrument and the varying response of V1298\,Tau’s activity in the blue and red wavelengths -- due to the different contrast of active regions at different wavelengths -- introduced an additional complexity to the analysis of the RVs. We approached this issue allowing different amplitudes in the stellar model for the instruments with different spectral ranges. To better monitor the changes caused by stellar activity, we performed contemporaneous $V$-band photometry with a cadence of 1 observation every 8 hours using the Las Cumbres Observatory (LCOGT) network\cite{Brown2013LCO}. We obtained 250 measurements with a typical precision of 10 parts per thousand (ppt). The data showed variations of up to 60 ppt, almost twice as large as those found in the K2 data. This is not surprising for a spot-dominated photosphere, as the Kepler passband is more extended towards red wavelengths, where the spot contrast is smaller.

  \noindent To mitigate the effects of the stellar activity on the RV data of V1298 Tau, we opted for a global model combining the K2 photometry, RV, and one contemporaneous activity proxy, which in our case is the LCOGT photometry. We relied on a Gaussian processes (GP) regression\cite{Rasmussen2006}, that benefits from an adequate observing cadence, like the one provided by the LCOGT data, to account for the stellar activity contribution.  The model uses the K2 photometry to constrain the planetary orbital periods, phases and radii. To take advantage of the photometric follow-up, we shared several hyperparameters between the RV and LCOGT time series. We used the LCOGT photometry and the RVs to constrain the timescales and amplitudes of the activity variations during our observing campaign, and the RVs to measure the masses of the planets. We added four Keplerian components to the RV model, representing the planets four known transiting planets.

\begin{figure}
\centering
        \includegraphics[width=\textwidth]{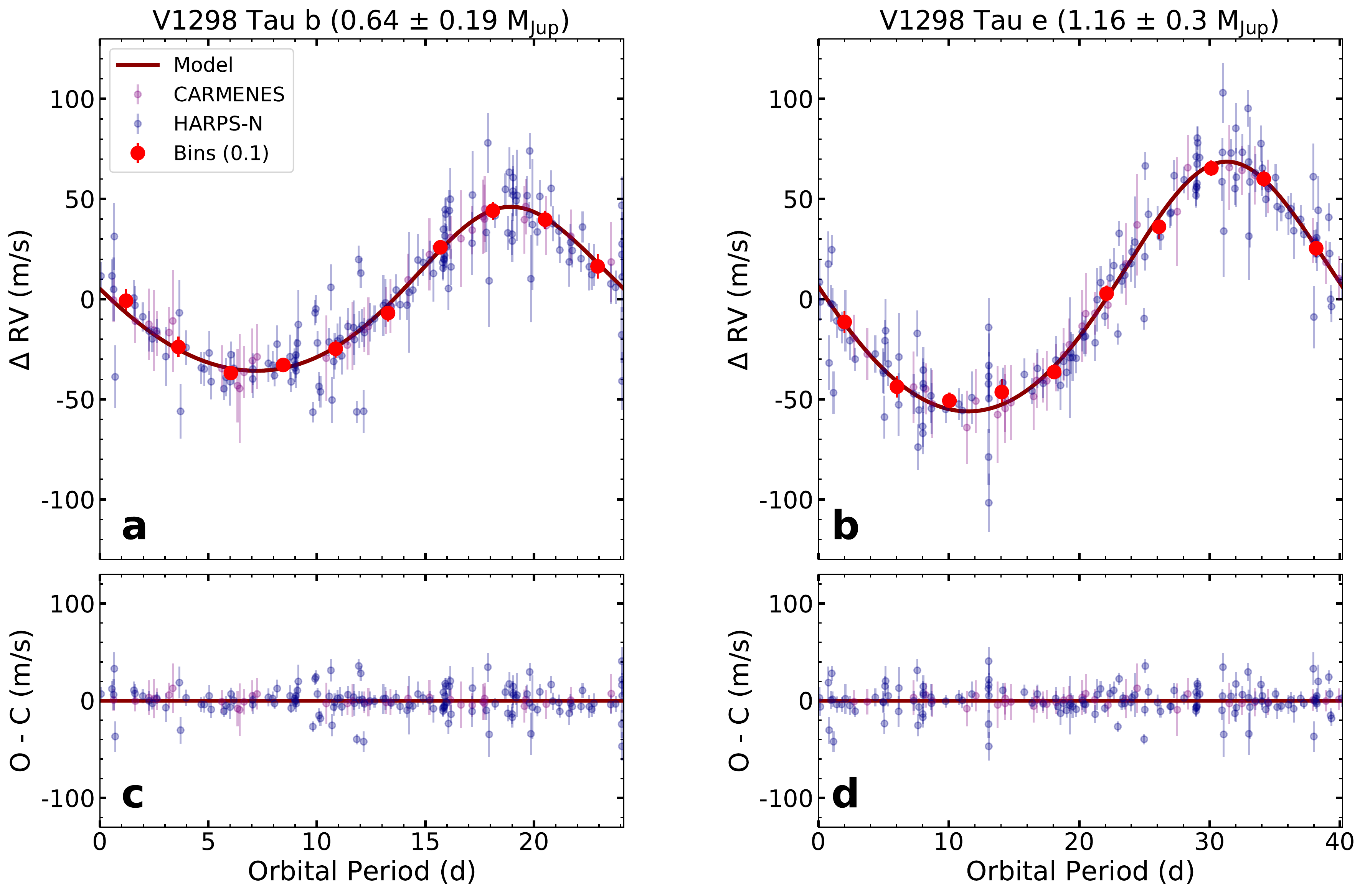}
        \caption{\textbf{Phase-folded plots of the RV signals for the two planets of the V1298\,Tau planetary system with significant mass measurements.} \textit{a}: Phase-folded representation of the best-fitting Keplerian orbit (red line) for V1298\,Tau b. \textit{b}: Same for V1298\,Tau e.  \textit{c and d}: Residuals after the fit for both cases. For a better visualisation, only HARPS and CARMENES data have been included. The large red dots show the data binned every 1/10th of the orbit. In all cases, 1$\sigma$ error bars (internal RV uncertainties) of the measurements are shown.}
        \label{rv_fold}
\end{figure}

 \noindent We obtained significant measurements of the RV semi-amplitudes induced by planet b of 41 $\pm$ 12 m~s$^{-1}$, and by planet e of 62 $\pm$ 16 m~s$^{-1}$. For the two inner-most planets, c and d, we can only set upper limits of 22 m~s$^{-1}$ and 25 m~s$^{-1}$ with a 99.7\% confidence, respectively. Planet e was originally detected with a single transit, insufficient for the accurate measurement of its orbital period\cite{DavidT2019}. Using the RV data, we obtained the detection at a period of 40.2 $\pm$ 1.0 d, on the short end of the range expected from the transit duration\cite{David19}. If this period is correct, the K2 mission missed transits right before and after its campaign by just a few days. We measured the orbital eccentricities for planets b and e to be of 0.13 $\pm$ 0.07 and 0.10 $\pm$ 0.09 respectively. Figure~\ref{rv_fold} shows the phase-folded curves of the RV signals attributed to planets b and e. The analysis of the RV and photometry during the 2019--2020 campaign yielded a stellar rotation period of 2.9104 $\pm$ 0.0019  days, with a semi-amplitude in RV of about 250 m~s$^{-1}$, and of 5\% of the flux  in the $V$-band photometry. Our analysis of the K2 data yielded orbital periods, times of transit, and relative radii consistent with the discovery paper\cite{David2019}. Table~S1 in the methods section shows the complete summary of our results, including all the alternative methods that we tested. We note that the determination of the period of planet e comes almost exclusively from the spectroscopic analysis. The new TESS observations of V1298\,Tau (Sectors 43 and 44, September -- October 2021) will provide an opportunity to detect a new transit of the planet, further solidifying its orbital period.

\begin{figure}

\includegraphics[width=16cm]{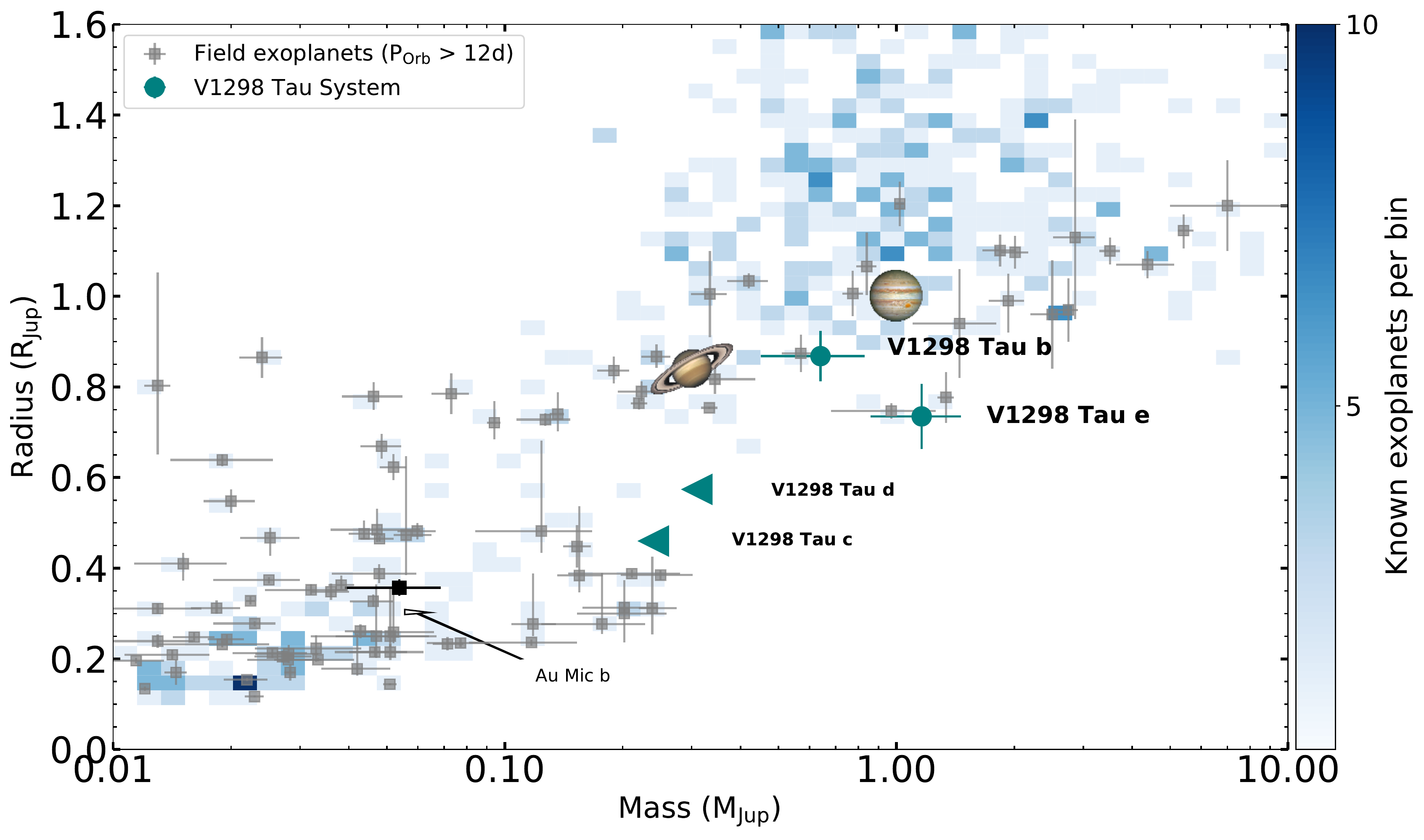}

\caption{\textbf{Planets of V1298\,Tau in the context of the known planets.} Histogram of the masses and radii of known planets for which the two parameters are determined with a precision better than 33\%. The planets orbiting V1298\,Tau are highlighted in teal symbols, with their 1$\sigma$ error bars. The left-pointing arrows show the upper limits for the masses of V1298\,Tau c and d. Jupiter and Saturn have been added for comparison. The planets at orbital periods longer than 12 days are shown with dark-grey symbols.} \label{mass_rad}
\end{figure}

 \noindent We derived the masses of the planets b and e to be 0.64 $\pm$ 0.19 M$_{\rm Jup}$ and 1.16 $\pm$ 0.30 M$_{\rm Jup}$, respectively, that, together with their orbital distances and the star's effective temperature, place them both in the category of warm Jupiters. For the same planets we derived radii of 0.868 $\pm$ 0.056 R$_{\rm Jup}$ and 0.735 $\pm$ 0.072 R$_{\rm Jup}$, respectively, compatible with previous measurements~\cite{David2019}. We derived densities of 1.20 $\pm$ 0.45 g cm$^{-3}$ and 3.6 $\pm$ 1.6 g cm$^{-3}$, respectively. V1298\,Tau b occupies a position in the mass-radius diagram compatible with the old giant planets of the Solar System (Figure~\ref{mass_rad}). Planet e is more compact and lies in a less populated region of the mass-radius diagram, resembling dense giant planets like HATS-17 b\cite{Brahm2016} and Kepler-539 b\cite{Mancini2016}. For the two smaller planets, c and d, we calculated 3$\sigma$ upper limits on their masses of 0.24 M$_{\rm Jup}$ and 0.31 M$_{\rm Jup}$, respectively, which set upper limits to the densities of 3.5 g cm$^{-3}$ and 2.4 g cm$^{-3}$, respectively. The combination of the masses of all the pairs in the system are Hill-stable.  Table~\ref{planet_parameters_1} shows the final planetary parameters adopted for the system.

\begin{figure}
\includegraphics[width=16cm]{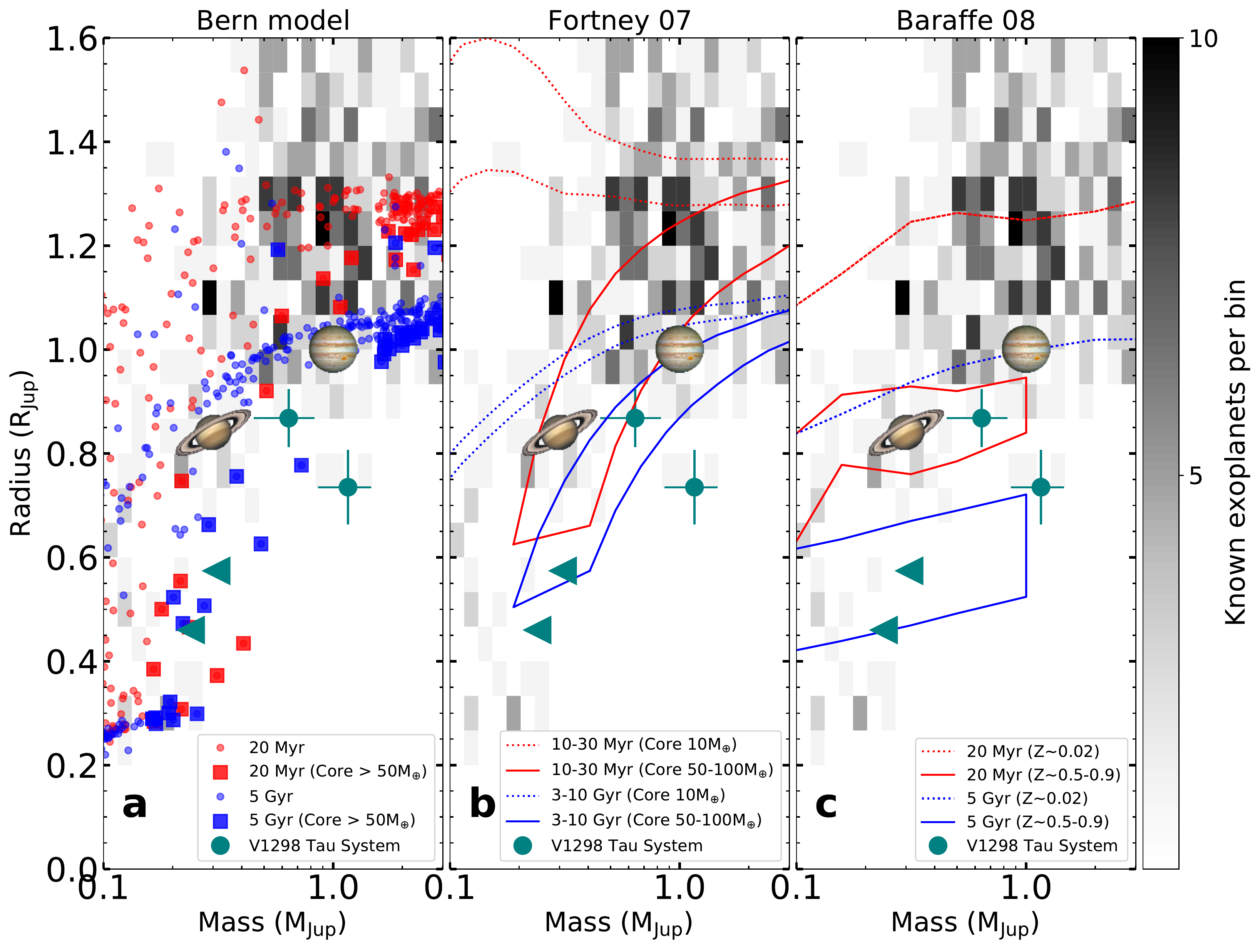}
\caption{\textbf{Planets of V1298\,Tau in the context of the models of planetary evolution.} \textit{a}: Masses and radii of the planets orbiting V1298\,Tau, with their 1$\sigma$ error bars, compared to the expected planetary population orbiting stars with solar mass and metallicity at 20 Myr and 5 Gyr\cite{Emsenhuber2020}. Core-heavy planets have been highlighted. \textit{b}: Masses and radii of the planets orbiting V1298 Tau compared to the mass radius-tracks from Fortney et al. for planets with light and heavy-cores (50 -- 100 M$_{\oplus}$)\cite{Fortney2007}. \textit{c}: Masses and radii of the planets orbiting V1298\,Tau compared to the mass radius-tracks by Baraffe et al. 2008 for planets with light and enriched cores (Z$\sim$ 0.5 - 0.9)\cite{Baraffe2008}. The left-pointing arrows show the upper limits for the masses of V1298\,Tau c and d. All panels show the histogram of the known population of exoplanets, and the positions of Jupiter and Saturn, as reference points.} \label{mass_rad_2}
\end{figure}

 \noindent Core accretion models of planetary evolution predict planets of $\sim$ 20 Myr to be at the early stages of their contraction phase, showing very large radii and low densities\cite{Fortney2007, Baraffe2008, Mordasini2012}. Our results indicate that V1298\,Tau b and e deviate from this picture. Figure~\ref{mass_rad} shows the position of the planets orbiting V1298\,Tau in a mass-radius diagram compared to the known population of exoplanets. Similar to the case of AU Mic b~\cite{Klein2021} -- the only other exoplanet of similar age with a mass measurement -- the mass-radius relation of the planets orbiting V1298\,Tau resembles that of the planets of our Solar System and of the general population of known transiting exoplanets. However, in contrast to the case of AU Mic b, the planets V1298\,Tau b and e seem incompatible with the expected population derived from these models of evolution of planetary systems.  Figure~\ref{mass_rad_2} shows the planets orbiting V1298\,Tau with the expected population of exoplanets orbiting 1 M$_\odot$ stars at the ages of 20 Myr and 5 Gyr (simulation NG76\cite{Emsenhuber2020} using the Bern model\cite{Mordasini2012}) and with the mass-radius tracks coming from other different models\cite{Fortney2007, Baraffe2008}. According to current theories these planets cannot reach this mass-radius configuration until hundreds of Myr later. Considering their densities, it is not expected that the planets orbiting V1298\,Tau will contract significantly in the future due to evaporation\cite{Poppenhaeger2021}. Our result suggests that some giant planets reach a mass-radius configuration compatible with the known mature population of exoplanets during their first 20 $\pm$ 10 Myr of age.
 
 \noindent An alternative explanation to the characteristics of V1298\,Tau b and e could be offered by an extreme enrichment in heavy elements compared to giant planets of the Solar System and the general population of transiting exoplanets\cite{Thorngren2016}. A fraction of heavy elements of 40--60\% and 60--80\% of the mass of planets b and e would partially reconcile our results with the core-accretion evolutionary models\cite{Fortney2007, Baraffe2008, Mordasini2012}. This enrichment would correspond to 3-20 times the fraction of heavy elements of Jupiter for V1298\,Tau b, and 5-25 times for V1298\,Tau e. Such high metallicities are not expected for planets orbiting stars with solar metallicity. Figure~\ref{mass_rad_2} shows the planets orbiting V1298\,Tau with the mass-radius tracks derived for core-heavy exoplanets at the ages of 20 $\pm$ 10 Myr and the age of the Solar System. The models for core-heavy exoplanets predict the planets will keep contracting, moving them to a region of the parameter space in which there are no known field exoplanets. If giant planets with such high metallicities existed, and this scenario were correct, it would suggest that the two planets formed further away from their host star, possibly at separations of several astronomical units, consistent with those observed in protoplanetary disks imaged by ALMA, and experienced large-scale migration and accretion of planetesimals\cite{Turrini2021}. Previous studies have shown that Neptune-sized exoplanets can be found in short orbits at the age of 10 Myr\cite{David16}. The original discovery of V1298\,Tau b\cite{DavidT2019} also pointed in the same direction for planets with Jupiter radii. The V1298 Tau system provides strong evidence that planets with masses similar to Jupiter can reach close-in orbits within the first 20 $\pm$ 10 Myr.



\begin{addendum}
 \item 
ASM acknowledges financial support from the Spanish Ministry of Science and Innovation (MICINN) under the 2019 Juan de la Cierva Programme.  J.I.G.H. acknowledges financial support from Spanish MICINN under the 2013 Ram\'on y Cajal program RYC-2013-14875.  A.S.M., J.I.G.H., R.R., C.A.P., N.L, M.R.Z.O., E.G.A., J.A.C., P.J.A., and I.R. acknowledge financial support from the Spanish Ministry of Science and Innovation through projects AYA2017-86389-P, PID2019-109522GB-C53, PID2019-109522GB-C51, AYA2016-79425-C3-3-P, PID2019-109522GB-C52, and PGC2018-098153-B-C33.M.D. acknowledges financial support from the FP7-SPACE Project ETAEARTH (GA No. 313014).  A.M., D.L., G.M., A.S. and S.D. acknowledge partial contribution from the agreement ASI-INAF n.2018-16-HH.0. S.B., D.L. and G.M, acknowledge partial contribution from the agreement ASI-INAF n.2021-5-HH.0. P.J.A acknowledges  financial support from the project SEV-2017-0709. SD, VdO, SB acknowledge support from  the PRIN-INAF 2019 "Planetary systems at young ages (PLATEA). DA thanksthe Leverhulme Trust for financial support.  I.R. acknowledges the support of the Generalitat de Catalunya/CERCA programme.  E.G.A acknowledges support from the Spanish State Research Agency (AEI) Project  No. MDM-2017-0737 Unidad de Excelencia “María de Maeztu”- Centro de Astrobiología (CAB, CSIC/INTA). Based on observations made with the Italian Telescopio Nazionale Galileo (TNG) operated by the Fundación Galileo Galilei (FGG) of the Istituto Nazionale di Astrofisica (INAF) at the Observatorio del Roque de los Muchachos (La Palma, Canary Islands, Spain). CARMENES is an instrument at the Centro Astron\'omico Hispano-Alem\'an
(CAHA) at Calar Alto (Almer\'{\i}a, Spain), operated jointly by the Junta de Andaluc\'ia and the Instituto de Astrof\'isica de Andaluc\'ia (CSIC). Based on data obtained with the STELLA robotic telescopes in Tenerife, an AIP facility jointly operated by AIP and IAC. This work makes use of observations from the LCOGT network. Based on observations made with the Nordic Optical Telescope, owned in collaboration by the University of Turku and Aarhus University, and operated jointly by Aarhus University, the University of Turku and the University of Oslo, representing Denmark, Finland and Norway, the University of Iceland and Stockholm University at the Observatorio del Roque de los Muchachos, La Palma, Spain, of the Instituto de Astrofisica de Canarias.

 \item[Author Contribution] A.S.M. wrote the main text of the manuscript. A.S.M, V.J.S.B., J.I.G.H, M.R.Z.O and C.d.B. wrote the methods section of the manuscript. A.S.M. and M.D. performed the radial velocity analysis. N.L., A.S., V. J. S. B., G. M., R. R., S. B., C.C.G., A.S.M., M.D., P. J. A. and M. W. coordinated the acquisition of the radial velocities. V. J. S. B., F.M., E.P., H.P. and E.E.B coordinated the acquisition of the photometry. A.S.M., B.T.P., F.F.B and T.G. performed the extraction of radial velocities. F.M. performed the extraction of the photometry. V.J.S.B., M.R.Z.O., J.I.G.H., C.d.B., H.M.T., D.S.A., N.L. and E.L.M. determined the stellar properties of V1298\,Tau and HD 284154. R.C., A.M., D.T. contributed to the discussion on planetary evolution. R.R.L, A.S., M.R.Z.O., V.J.S.B and G.L. organised the collaboration between the different teams. M.D., A.S., S.B., G.M., S.D., R.C., L.M., V.D., D.L., F.M., D.T., A.M. are members of the GAPS consortium. P.J.A., J.A.C, A.Q., A.R., I.R, M.R.Z.O, V.J.S.B., J.I.G, H., N.L and R.R.L are members of the CARMENES consortium. L.M. participated in the discussion of stellar activity. T.G., K.G.S. and M.W. are members of the STELLA consortium. All authors were given the opportunity to review the results and comment on the manuscript.

\item[Author Information] The authors declare that they have no
competing financial interests. Correspondence and requests for materials
should be addressed to A.S.M.~(email: asm@iac.es).
\end{addendum}


\begin{table*}
\fontsize{9}{11}\selectfont
\begin{center}
\caption{Stellar parameters of V1298\,Tau and its wide companion HD\,284154, with their 1$\sigma$ uncertainties} \label{stellar_params}
\begin{tabular}{ l c c l}
\\ \hline 
Parameter & V1298\,Tau & HD\,284154 & Reference\\ \hline
RA [h m s] & 04 05 19.59 & 04 05 14.35 &\cite{Gaia2018}\\
Dec [$^o$ $'$ $''$] & $+$20 09 25.6 & $+$20 08 21.5 & \cite{Gaia2018}\\
$\mu_{\alpha} \ {\rm cos} \delta$ & 5.23 $\pm$ 0.13 & 5.04 $\pm$ 0.12 & \cite{Gaia2018} \\
$\mu_{\delta}$ [mas\,a$^{-1}$] &  -16.08 $\pm$ 0.05 & -16.32 $\pm$ 0.05 & \cite{Gaia2018}\\
Parallax [mas] & 9.214 $\pm$ 0.05 & 9.202 $\pm$ 0.060 &\cite{Gaia2018}\\
Distance [pc] & 108.5 $\pm$ 0.7 & 108.7 $\pm$ 0.7 &\cite{Gaia2018}\\
Spectral type & K1 & G2 &\cite{Nguyen2012, Nesterov1995}\\ 
$V$ [mag] & 10.12 $\pm$ 0.05 & 8.51 $\pm$ 0.02 &\cite{Hog2000}\\
$G$ [mag] & 10.0702 $\pm$ 0.0007 & 8.3561 $\pm$ 0.0005 &\cite{Gaia2018} \\
$J$ [mag] &  8.687 $\pm$ 0.023 & 7.287 $\pm$ 0.020 &\cite{Skrutskie2006} \\
$K$ [mag] & 8.094 $\pm$ 0.021 & 6.947 $\pm$ 0.026 &\cite{Skrutskie2006} \\
$U$ [km~s$^{-1}$] & $-$12.63 $\pm$ 0.03 &  -12.90 $\pm$ 0.64     & This work \\
$V$ [km~s$^{-1}$] & $-$6.32 $\pm$ 0.06 & -6.32 $\pm$ 0.10  & This work \\
$W$ [km~s$^{-1}$] & $-$9.1 9$\pm$ 0.06 & -9.49 $\pm$ 0.28  & This work \\
Age [Myr] & 20 $\pm$ 10 & 20 $\pm$ 10 & This work \\
Luminosity [L$_{\odot}$]  &  0.954 $\pm$ 0.040   & 4.138 $\pm$ 0.040$^{1}$ & This work \\
Effective temperature [K]  & 5050 $\pm$ 100  & 5700 $\pm$ 100 & This work \\
Mass [M$_\odot$]  & 1.170 $\pm$ 0.060 & 1.28 $\pm$ 0.06$^{2}$ & This work \\
Radius [R$_\odot$] & 1.278 $\pm$ 0.070  & 1.477 $\pm$ 0.082$^{2}$ & This work  \\
Rotation period [d] & 2.91 $\pm$ 0.05 & ... & This work \\
$v$ ~sin $i$ [km~s$^{-1}$]  & 23.8 $\pm$ 0.5 & 10 $\pm$ 1 & This work \\
$~\rm[Fe/H]$ [dex] & 0.10 $\pm$ 0.15 & 0.05 $\pm$ 0.15 & This work \\
\hline 

\end{tabular}

\end{center}
1: Corresponds to the binary system. \\
2: Measurement for each individual component assuming an equal-mass binary.
\end{table*}

\begin{table*}
\fontsize{9}{11}\selectfont
\begin{center}
\caption{Planetary parameters for the V1298\,Tau system.} Parameters show 1$\sigma$ uncertainties. Upper and lower limits show the 99.7$\%$  confidence interval. T0 given in BJD - 2450000. \label{planet_parameters_1} 
\begin{tabular}{ l c c c c c}
\\ \hline 
Parameter & V1298\,Tau b & V1298\,Tau c & V1298\,Tau d & V1298\,Tau e\\ \hline
{\it P}$_{\rm orb}$ [d] & 24.1399 $\pm$ 0.0015 & 8.24892 $\pm$ 0.00083 & 12.4058 $\pm$ 0.0018 & 40.2 $\pm$ 1.0 \\
{\it T}$_{\rm 0}$ [d] &  7067.0486$\pm$ 0.0015 & 7064.2801$\pm$ 0.0041 & 7072.3907$\pm$ 0.0063 &7096.6226$\pm$ 0.0031 \\
{\it a} [au] & 0.1719 $\pm$ 0.0027 & 0.0841$\pm$ 0.0013 & 0.1103 $\pm$ 0.0017 & 0.2409 $\pm$ 0.0083\\
{\it R}$_{\rm p}$/{\it R}$_{*}$ & 0.0698$\pm$ 0.0024 &  0.0371$\pm$ 0.0019 &  0.0464$\pm$ 0.0020 & 0.0583$\pm$ 0.0040 \\
{\it R}$_{\rm p}$ [R$_{\rm Jup}$] & 0.868 $\pm$ 0.056 & 0.460$\pm$ 0.034 & 0.574 $\pm$ 0.041 & 0.735 $\pm$ 0.072 \\
Incl. [Deg] &  \textgreater~ 88.7 &  \textgreater~ 87.5 & \textgreater~  88.3 & \textgreater~ 89.0\\
{\it K}$_{\rm RV}$ [m~s$^{-1}$] &  41 $\pm$ 12 & \textless~ 22 & \textless~ 25 & 62 $\pm$ 16 \\
{\it e} & 0.134 $\pm$ 0.075 & \textless~0.30& \textless~0.20 & 0.10 $\pm$ 0.091\\
{\it m} [M$_{\rm Jup}$] & 0.64 $\pm$ 0.19 & \textless~0.24  & \textless~0.31 & 1.16 $\pm$ 0.30\\
$\rho$ [g cm$^{-3}$] & 1.20 $\pm$ 0.45 & \textless~ 3.5 & \textless~2.4 & 3.6 $\pm$ 1.6 \\

\hline

\end{tabular}
\end{center}
\end{table*}

\newpage

\newenvironment{methods}{%
    \section*{Methods}%
    \setlength{\parskip}{12pt}%
    }{}

\begin{methods}

\section{DATA}\label{sect_data}

\subsection{HARPS-N and CARMENES RVs.} 

HARPS-N\cite{Cosentino2012} is a fibre-fed high resolution échelle spectrograph installed at the 3.6 m Telescopio Nazionale Galileo of the Roque de los Muchachos Observatory (La Palma, Spain). It has a resolving power of  115 000 over a spectral range of 360$-$690 nm and is contained in a temperature- and pressure-controlled vacuum vessel to avoid spectral drifts due to temperature and air pressure variations. It is equipped with its own pipeline, providing extracted and wavelength-calibrated spectra, as well as RV measurements and other data products, such as cross-correlation functions and the bisector of their line profiles. We obtained 132 observations between 2019 and 2020: 72 of those measurements under Spanish time and the remaining 60 in the context of the GAPS programme\cite{Covino2013GAPS,Carleo2020}, a long-term, multi-purpose, observational programme aimed to characterise the global architectural properties of exoplanetary systems. On-source integration times were typically set to 900 -- 1200 s.  Using the HARPS-N data, we obtained the $S_{\rm MW}$\cite{Noyes1984}, H${\alpha}$\cite{GomesdaSilva2011}, Na I\cite{Diaz2007} and TiO\cite{Azizi2018} chromospheric indicators. 

\noindent The CARMENES instrument\cite{Quirrenbach2014} consists of a visual (VIS) and near-infrared (NIR) spectrographs covering 520 -- 960 nm and 960 -- 1710 nm with a spectral resolution of 94 600 and 80 400, respectively$^{19}$. It is located at the 3.5 m Zeiss telescope at the Centro Astronómico Hispano Alemán (Almería, Spain).  We extracted the spectra with the \texttt{CARACAL} pipeline, based on flat-relative optimal extraction\cite{Zechmeister2014}. The wavelength calibration that was performed by combining hollow cathode lamps (U-Ar, U-Ne, and Th-Ne) and Fabry-Pérot etalons. The instrument drift during the nights is tracked with the Fabry-Pérot in the simultaneous calibration fibre. We obtained 35 observations between 2019 and 2020.

\noindent RVs for HARPS-N and CARMENES were obtained using  \texttt{SERVAL}. This software builds a high signal-to-noise template by co-adding all the existing observations, and then performs a maximum likelihood fit of each observed spectrum against the template, yielding a measure of the Doppler shift and its uncertainty. We obtained typical RV precisions of 9 and 15 m~s$^{-1}$ for HARPS and CARMENES VIS measurements. For CARMENES NIR measurements we obtained a typical RV precision of 55 m~s$^{-1}$. We measured an RMS of the RVs of 260 m~s$^{-1}$, 197 m~s$^{-1}$, and 195 m~s$^{-1}$ for the HARPS-N, CARMENES VIS, and CARMENES NIR respectively. Contrary to what was expected, the CARMENES NIR RVs showed no significant reduction in dispersion compared to the VIS RVs. These NIR data show a large difference in precision compared to the visible arm and adds no new temporal information. We opted to avoid increasing the complexity of the model and relied exclusively on the RVs coming from the visible arm.

\subsection{HERMES RVs.}

The High-Efficiency and high-Resolution Mercator Échelle Spectrograph (HERMES)\cite{Raskin2011} is installed at the 1.2m Mercator telescope at the Roque de los Muchachos Observatory (La Palma, Spain). The spectra were automatically processed by the HERMES pipeline,  but later we derived our own RV measurements by cross-correlating the spectra with a K1 synthetic template \cite{Baranne1996} taken from the HARPS-N reduction pipeline. This made the effective wavelength range used for the RV calculation similar to the HARPS-N wavelength range. We obtained 35 measurements during 18 individual winter nights in 2019 -- 2020, with the goal of monitoring the shape of the activity variations. We obtained a typical RV precision of 55 m~s$^{-1}$ per observation and an RMS of the RVs of 294 m~s$^{-1}$.

\subsection{SES RVs.}

The STELLA échelle spectrograph (SES)\cite{Strassmeier2004} is a high resolution spectrograph installed at the 1.2m STELLA telescope at the Teide Observatory (Tenerife, Spain) . It has a resolving power of 55 000 over a wavelength range of 390 -- 870 nm$^{20}$. RVs were obtained by the automatic reduction pipeline, by cross-correlating the spectra with a synthetic stellar template \cite{Baranne1996} with a temperature of 5000K. We obtained 61 epochs  spread across 3 months, during the winter of 2019--2020, with the goal of following the activity variations of V1298 Tau.  We obtained a typical RV precision of 117 m~s$^{-1}$ per observation and an RMS of the RVs of 309 m~s$^{-1}$.

\subsection{LCOGT $V$-band photometry.}

We observed V1298 Tau with the 40 cm telescopes of the Las Cumbres Observatory (LCOGT)$^{25}$. The 40 cm telescopes are equipped with a 3k$\times$2k SBIG CCD camera with a pixel scale of 0.571\,arcsec providing a field of view of 29.2$\times$19.5\,arcmin.

\noindent We observed V1298Tau in the $V$-band every 8h over 4 months, during the winter of 2019--2020. The raw images were reduced by LCO's pipeline \texttt{BANZAI} and aperture photometry was performed on the calibrated images using \texttt{AstroImageJ}\cite{Collins2017}. For each night, we selected a fixed circular aperture in \texttt{AstroImageJ} and performed aperture photometry on the target and 5 reference stars of similar brightness. We obtained 250 $V$-band measurements with a typical precision of 0.5\% in relative flux. 

\subsection{K2 photometry.} 

As complementary data to the spectroscopic dataset, we downloaded the available photometric light curve obtained by the Kepler Space Telescope \textit{K2} mission $^{13}$ from the  Mikulski Archive for Space Telescopes (MAST). This photometric dataset was taken in the long cadence mode, characterised by 30-min integration time. We adopted the EVEREST 2.0\cite{Luger2018} light curve, which corrects for K2 systematics using a variant of the pixel-level decorrelation method\cite{Deming2013}. This time-series covers a time-span of about 71d (one Kepler quarter) from 2015 February 8 to 2015 April 20 , which corresponds to the K2 Campaign 4.

\subsection{FIES spectroscopy.}

We took a single exposure of 2100s of HD~284154 with the 
high-resolution ($R$~67 000) FIES spectrograph at the 2.6m-NOT telescope 
of the Roque de los Muchachos Observatory (La Palma, Spain). Observations made on 18 August 2020.

\section{ Stellar parameters of V1298 Tau}

\subsection{Membership to Group 29.}
V1298 Tau is the low-mass companion of the warmer, G0-type star HD 284154 at a projected separation of 97.7 arcsec (or 10600 AU at the distance of the system). The pair belongs to the recently identified Group 29\cite{Oh2017}, which is a young, sparse association of coeval stars in the Taurus region, all of which share very similar proper motions and distances based on the Tycho-Gaia astrometric catalog (TGAS)\cite{Luhman2018,Andrews2017}. Using updated trigonometric parallaxes and other astrometric determinations provided by the Gaia Data Release 2$^{26}$, we confirm that both V1298 Tau and HD 284154 are proper motion companions located at a distance that is compatible with that of the Group (see table 1). The Galactic velocities $U, V, W$ of V1298 Tau and HD 284154 completely overlap with the distribution of the space motions of Group 29 members, thus providing additional support to the membership of V1298 Tau and HD 284154 in this association.

\subsection{Effective temperature, surface gravity and metallicity of V1298 Tau.}

We derived the stellar parameters and metallicity of V1298 Tau using the 
high-resolution HARPS-N spectra. The extracted, blaze-corrected 2D spectra were corrected for barycentric velocity  (varying from $+$10.8 to $-30.4$~\kmso) and for RV (varying from  $+$14.4 to $+$15.7~\kmso) and normalised to unity order by order with a third-order polynomial using our own IDL-based automated code\cite{Jonay2020}. All orders of all spectra were co-added and merged with a wavelength step of  0.01 {\AA} per pixel. The resulting 1D spectrum shows a signal-to-noise ratio of $\sim$~107, 222, 284, 412 and 391 at 4200, 4800, 5400, 6000, and 6600~{\AA}, respectively.

\noindent Using the same automated code, we normalised and combined the HARPS-N spectra 
from our RoPES RV program\cite{Masca2018} of three other stars (HD\,220256, HD\,20165 and $\epsilon$~Eri) with similar spectral types to be used as comparison stars. We compared the HARPS-N spectra of these templates with that of V1298 Tau to derive the projected rotation velocity of V1298 Tau. To reduce the computing load we restricted the calculation to the spectral range 5355--5525~{\AA} and performed a Markov Chain Monte Carlo (MCMC) simulation with 5000 chains implemented in \texttt{ emcee}\cite{Foreman-Mackey2013}. The mean stellar projected rotation velocity of V1298 Tau obtained from the three templates is $23.8\pm0.5$ km$s^{-1}$, which is consistent with the measurement derived from the stellar radius and rotation period ($22.2\pm1.3$ km$s^{-1}$).

\noindent To estimate the stellar parameters of V1298 Tau ($T_{\rm eff}$ and $\log g$, and metallicity, [Fe/H]\footnote{$A(X) = log[N({\rm X})/N({\rm H})] + 12$ and [Fe/H] = $A({\rm X}) - A_\odot({\rm X})$ with ${\rm X} = {\rm Fe}$}), we used three different codes, which allowed us to check the consistency of the results. First, we used the {\ferre} code\cite{Callende2006} with a grid of synthetic spectra\cite{Callende2018} with a micro-turbulence velocity fixed at $\xi_{\rm mic} = 1.5$~\kms to fit the HARPS-N spectrum of V1298 Tau, providing $T_{\rm eff}/\log g/$A(Fe)~$=5010/4.48/7.20$ (note that the canonical solar Fe abundance is $A_\odot({\rm Fe})=7.50$\cite{Asplund2009}). For comparison, we also analysed the HARPS-N spectrum of the star $\epsilon$~Eri and the Kurucz solar ATLAS spectrum\cite{Kurucz1984}, both broadened with a rotation profile of 24~\kmso, obtaining $T_{\rm eff}/\log g/$A(Fe)~$=5085/4.91/7.24$ and~$=5912/4.74/7.31$, respectively. {\ferre} uses a running mean filter to normalise both synthetic and observed spectra and fits a wide spectral range of the HARPS-N spectrum (4500--6800~{\AA}), masking out the Balmer and NaID lines, which for the young V1298 Tau star could show their cores in emission. Taking the analysis of the Solar ATLAS as the solar reference, this method \#1 gives [Fe/H]~$=-0.11$ as the metallicity of V1298 Tau. We suspect that slightly low metallicity is related to the relatively high microturbulence adopted for the grid of synthetic spectra\cite{Callende2018}. The expected microturbulent velocity should be $\xi_{\rm mic} = 0.85$~\kmso~\cite{Dutra-Ferreira2016}, so this may be the reason why the derived metallicity with this method appears to be slightly lower than the solar value. 

\noindent Second, we used the \texttt{SteParSyn} code (Tabernero et al. 2021, in preparation), 
a Bayesian code that uses a synthetic grid of small spectral regions of 3~{\AA} 
around 95 Fe lines with a fixed $\xi_{\rm mic} = 0.85$~\kmso. The result of this second method is $5041/4.24/7.62$ a metallicity  [Fe/H]~$=+0.16$.

\noindent Third, we implemented a Bayesian python code that compares the observed spectrum with a synthetic spectrum in the spectral range 5350 -- 5850\AA{} (see Figure~\ref{syn_spec}).
We performed a Markov Chain Monte Carlo (MCMC) simulation with 5000 chains implemented in \texttt{emcee}\cite{Foreman-Mackey2013}. We used a small 3x3x3 grid of synthetic spectra with values 
$T_{\rm eff}/\log g/$A(Fe) of $4750 - 5250 / 3.5 - 4.5 / 7.0 - 8.0$ and steps of 
250~K / 0.5~dex / 0.5~dex, computed with the {\sc SYNPLE} code, assuming a micro-turbulence $\xi_{\rm mic}=0.85$~\kmso, and ATLAS9 model  atmospheres with solar $\alpha$-element abundances\cite{Castelli2003} ([$\alpha$/Fe]=0), and the same linelist as in method \#1. 
This method \#3 delivered $T_{\rm eff}/\log g/$A(Fe)~$=5071/4.25/7.44$ and a metallicity [Fe/H]~$=+0.07$ for V1289 Tau (the result for the broadened solar ATLAS is 5753/4.48/7.37).

\noindent Taking into account the slightly different results from the three different methods we adopted the values $T_{\rm eff}/\log g/$[Fe/H]~$=5050\pm100/4.25\pm0.20/+0.10\pm0.15$ for V1298 Tau. Finally, we checked the derived effective temperature 
by applying the implementation of the InfraRed Flux Method (IRFM\cite{Jonay2009}). Using the available photometry in the infrared bands JHK$_S$ from the Two Micron All-Sky Survey (2MASS$^{30}$) and the Johnson V magnitude from the AAVSO Photometric All-Sky Survey (APASS\cite{Henden2012}), and adopting $E(B-V)=0.061$ from the dust maps\cite{Schlegel1998} corrected\cite{Bonifacio2000} using the distance of 108~pc to V1298 Tau $^{40}$, 
we applied the IRFM to obtain a $T_{\rm IRFM}=5047\pm66$~K, in agreement with the spectroscopic value. Assuming an extinction $E(B-V)=0.024\pm0.015$ $^{11}$, 
we got $T_{\rm IRFM}=4947\pm67$~K, and $4928\pm67$~K for $E(B-V)=0$.

\subsection{Effective temperature, surface gravity and metallicity of HD 284154.}

HD~284154 is resolved as a double-lined spectroscopic binary (see Figure~\ref{spec_binary}), which has been identified as a wide binary of V1298 Tau\cite{Andrews2017}. Using the FIES spectrum, we estimated a RV difference of $\delta RV= 43.6\pm1.0$~\kms between the two stellar components, an identical stellar rotation of $V_{\rm rot}= 10\pm1$~\kms derived from the double-peaked cross-correlation function (CCF) of the observed FIES spectrum cross-correlated with a mask of isolated and relatively strong Fe lines. Assuming both binary components have the same mass (see below), the center-of-mass RV is $+14.8\pm0.7$ \kmso, perfectly compatible with the center-of-mass RV of V1298 Tau. 

\noindent We applied method \#3 to derive the stellar parameters and metallicity of both
stellar components of HD~284154. We assumed both stars have the same luminosity and origin, and therefore the same stellar mass and metallicity. We then generated a grid of synthetic spectra in the parameter range $T_{\rm eff}/\log g/$A(Fe) of 
$5500 - 6000 / 3.5 - 4.5 / 7.0 - 8.0$ and steps of 250~K / 0.5~dex / 0.5~dex, 
with a fixed $\xi_{\rm mic} = 0.95$~\kmso, and adding the fluxes of each component separated by $\delta RV= 43.6$~\kms (see Figure~\ref{spec_binary}).
We obtained $T_{\rm eff}/\log g/$[Fe/H]~$=5700\pm100/4.35\pm0.20/+0.05\pm0.15$ for 
HD~284154.

\subsection{Lithium abundance.}
We used the MOOG code\cite{Sneden1973} and ATLAS9 model atmospheres to derive the lithium abundances of V1298 Tau and  HD~284154 (see Figure~\ref{spec_li}), with the approximation of local thermodynamical equilibrium (LTE). We applied the non-LTE corrections\cite{Lind2009} to get a Li abundance of A(Li)~$=3.43\pm0.15$ and $3.24\pm0.15$, for V1298 Tau and HD~284154, respectively, roughly consistent with the solar meteoritic value\cite{Asplund2009}.

\subsection{Masses, radii and luminosities.}

We determined the bolometric luminosity of both V1298 Tau and HD 284154 by transforming the observed magnitudes into bolometric magnitudes using {\sl Gaia} distances and colour-bolometric corrections\cite{Peacock13}. We confirmed that the obtained values (shown in Table~1) are fully compatible at the 1-$\sigma$ level with those derived from the integration of the photometric spectral energy distributions using the Virtual Observatory Spectral Energy Distribution Analyzer\cite{Bayo08} (VOSA) tool for the stellar effective temperatures. The radius of each star was then obtained from the Stefan-Boltzmann equation; we split the luminosity of HD 284154 in two identical parts to account for its nearly equal mass binary nature and arbitrarily augmented the error in the luminosity determination of each component by a factor of two. Masses were obtained from the comparison of the derived effective temperatures and bolometric luminosities with various stellar evolutionary models available in the literature\cite{Baraffe15,pisa,Bressan12}. All models are consistent within the error bars. Uncertainties in the mass determination account for the temperature and luminosity uncertainties and also for the dispersion of the results from the different models including models with slightly different metallic composition. We obtain mass and radius estimates of 1.170 $\pm$ 0.060  M$_{\odot}$ and  1.278 $\pm$ 0.070 M$_{\odot}$ for V1298 Tau, and 1.28 $\pm$ 0.06 R$_{\odot}$ and 1.477 $\pm$ 0.082 R$_{\odot}$ for each of the stars of HD 284154. All values are provided in Table~1. We additionally inferred the stellar parameters for V1298 Tau and HD 284154 from stellar evolution models using a Bayesian inference method\cite{Burgo18}. This Bayesian analysis makes use of the PARSEC v1.2S library of stellar evolution models\cite{Bressan12}. It takes the absolute G magnitude (using the parallax), and the colour G$_{BP}$-G$_{RP}$from $Gaia$ DR2$^{40}$, and assumes solar metallicity, with [Fe/H] = 0.00 $\pm$ 0.20, returning theoretical predictions for other stellar parameters. For HD\ 284154 it was assumed that this binary (whose Gaia photometry was corrected by adding 0.7526 mag) is in the pre-main sequence phase due to the expected youth of the association. For V1298 Tau we obtained a Log$~L$ of --0.040 $\pm$ 0.009, an effective temperature of 4929 $\pm$ 32 K, a mass of 1.17 $\pm$ 0.03 M$_\odot$, a radius of 1.310 $\pm$ 0.027 R$_\odot$ and a log$~g$ of 4.271 $\pm$ 0.028. For HD 284154 we obtained a Log$~L$ of 0.299 $\pm$ 0.009, an effective temperature of 5699 $\pm$ 55 K, a mass of 1.263 $\pm$ 0.013 M$_\odot$, a radius of 1.45 $\pm$ 0.04 R$_\odot$ and a log$~g$ of 4.218 $\pm$ 0.020.

\subsection{Age estimation.}\label{age_v1298}

Figure~\ref{gaia_colours} shows that the photometric sequence of Group 29 is compatible with the isochrones of 10-30 Myr. This photometric sequence of Group 29 is sub-luminous compared to that of the Upper Scorpius association, and is very similar to that of the Beta Pictoris moving group. This suggests that the Group 29 and, hence, the V1298 Tau system, is older than the Upper Scorpius association (5-11 Myr $^{10}$) and has a similar age to Beta Pictoris (20 $\pm$ 10 Myr\cite{Miret-Roig2020}). The rotation period of V1298 Tau (2.865 $\pm$ 0.012 d$^{9}$) points in a similar direction. It fits perfectly with stars with similar spectral types of very young associations such as Rho Ophiuchus, Taurus, Upper Scorpius and the Taurus foreground population\cite{Rebull2020}, with ages in the range of 1$-$30 Myr, but rotates faster than stars of similar spectral types in open clusters such as the Pleiades ($\sim$110 Myr\cite{Dahm15}) or Praesepe (600-800 Myr\cite{gossage18}). The lithium content of V1298 Tau also allows to constrain its age, since this element is destroyed in low-mass stars on timescales of tens of million years. Comparing the equivalent width of the lithium line of V1298 Tau (400 m\AA) with that of stars in open clusters and young moving groups of different ages\cite{GutierrezAlbarran2020}, we can conclude that the lithium content of V1298 Tau is compatible with an age of 1 -- 20 Myr, and is larger than in stars in open clusters such as IC\,2391 and IC\,2602, with estimated ages of 35 -- 50 Myr\cite{Barrado99}. V1298 Tau exhibits an X-ray emission of 4.58$^{+1.71}_{-1.44}$ 10$^{30}$ erg$~$s$^{-1}~$\cite{Wichmann1996},  compatible with a stronger activity than the stars in the Pleiades\cite{Micela1999,Fang2018}, and has a UV excess (NUV$-$J\,=\,8.15 $\pm$ 0.05 mag), characteristic of stars younger than $\sim$100 Myr\cite{Findeisen2011}. In addition, employing the same Bayesian inference method used in the previous section\cite{Burgo18}, we also obtained estimates for the ages of V1298 Tau and HD 284154 of 9 $\pm$ 2 Myr and 13 $\pm$ 4 Myr, respectively. Using all the previous results, we can constraint the age of V1298 Tau to be 20 $\pm$ 10 Myr.

\section{Modelling}\label{sect_model}

We fitted the K2 photometry, LCOGT photometry and RV time-series simultaneously, and modelled the activity signals in RV and the LCOGT photometry using Gaussian Processes (GP) with \texttt{celerite}\cite{Foreman-Mackey2017}. We used a variation of the quasi-periodic Kernel described in equation 56 of the original \texttt{celerite} article, with the explicit addition of a second mode at half the rotation period (PQP2 from now on): 

\begin{equation}
\fontsize{8}{11}\selectfont
 k(\tau) = {A^{2} \over{2+C}} \Bigg[ e^{-\tau/L_{1}}\Bigg(cos({{2\cdot \pi \tau}\over{P_{\rm rot}}}) + (1+C)\Bigg)    
 + {\Delta^{2}} \cdot e^{-\tau/L_{2}}\Bigg(cos({{4\cdot \pi \tau}\over{P_{\rm rot}}}) + (1+C)\Bigg) \Bigg] 
 + (\sigma^2 (t) + \sigma^2_{j}) \cdot \delta_{\tau}
\end{equation}

\noindent where \textit{A} represents the covariance amplitude, \textit{P}$_{\rm rot}$ is the rotation period, \textit{L$_{1}$} and \textit{L$_{2}$} represent the timescale of the coherence of the periodicity at the rotation period and its first harmonic, $\Delta$ represents the scaling in amplitude of the variability at the first harmonic of the rotation period, and \textit{C} the balance between the periodic and the non-periodic components. The equation also includes a term of uncorrelated noise ($\sigma$), independent for every instrument, added quadratically to the diagonal of the covariance matrix to account for all un-modelled noise components, such as uncorrected activity or instrumental instabilities. $\delta_{\tau}$ is the Kronecker delta function, and $\tau$ represents an interval between two measurements, $t-t'$. This Kernel behaves similarly to the classical quasi-periodic Kernel\cite{Haywood2014}. The  base version of this \texttt{celerite} Kernel was successfully used to to model the variations of Proxima Centauri to the level of the instrumental precision\cite{Masca2020}. To model the activity variations in the K2 photometry we used a combination of two simple harmonic oscillators (SHO) centred at the rotation period of the star and its first harmonic. This Kernel has been shown to appropriately model the photometric variations of V1298 Tau$^{9}$. The SHO Kernel is described in equation 2. To better constrain the behaviour of the GP in its description of the activity-induced RV variations, some of the hyper-parameters are shared between the GP of the LCOGT photometry and the RV $^{26}$. The period and timescales of coherence of the variability are shared parameters, while the amplitudes and mix factors are independent. As activity signals are known to have a chromatic dependence $^{15,83}$, we split the dataset by instruments and gave independent amplitudes of the activity signals to each instrument. The analysis considered a zero-point value and a noise term (jitter) for each dataset as free parameters to be optimised simultaneously, with the exception of the K2 data. For the K2 data we opted to manually include the white-noise component given in the discovery paper\cite{David19}. The K2 observations were obtained in 2017, while the LCOGT and RV data was obtained during 2019 and 2020. As the activity is not expected to remain stable after such a long time, we used two groups of hyper-parameters for the two different observing campaigns. We measure the final planetary parameters by fitting transits of the K2 lightcurve using the \texttt{pytransit} package\cite{Parviaine2015}, with quadratic limb darkening\cite{MandelAlgol2002}, and Keplerian orbits implemented with \texttt{Radvel}\cite{Fulton2018} in the RVs. 

\begin{equation}
\fontsize{8}{11}\selectfont
 k(\tau) = {A^{2}} e^{-\tau/L}  \left\{\begin{array}{cc}cosh(\nu {2 \pi \tau}/P_{rot})+{{P_{rot}}\over{2 \pi \nu L }}sinh(\nu {2 \pi \tau}/P_{rot}) , P_{rot} > 2 \pi L\\2 (1 + {{2 \pi \tau}\over{P_{rot}}}) , P_{rot} = 2\pi L\\cos(\nu {2 \pi \tau}/P_{rot}) + {{P_{rot}}\over{2 \pi \nu L}} sin(\nu {2 \pi \tau}/P_{rot}), P_{rot} < 2 \pi L\end{array}\right\}
 + (\sigma^2 (t) + \sigma^2_{j}) \cdot \delta_{\tau}
\end{equation}

\noindent with $\nu = (1 - (2L/P_{rot})^{-2})^{1/2}$.

\noindent To sample the posterior distribution and obtain the Bayesian evidence of the model (i.e. marginal likelihood, Ln$Z$) we relied on Nested Sampling \cite{Skilling2004} using \texttt{dynesty} \cite{Speagle2020}. We initialised a number of live points equal to $N \cdot (N+1) / 2$, with $N$ being the number of free parameters. 

\noindent We detected the signals corresponding to planets b and e (figure 1) and derived upper limits for the amplitudes of the signals corresponding to planets c and d (Figure~\ref{rv_fold_2}). This was our most significant model, with a measured Ln$Z$ of -- 4472. To confirm our results we repeated the analysis described above using the combination of two simple harmonic oscillators (SHO) to model the RV and LCOGT variations. We obtained a similar result, with larger amplitude for planet b and smaller uncertainties, but larger RMS of the residuals. This model proved to be less significant (Ln$Z$ of -- 4549). We also attempted to confirm the results using the Quasi-periodic GP Kernel to model the activity variations in the RV and LCOGT data, implemented using \texttt{George}\cite{Ambikasaran2015George} (Eq. 3). Previous studies have found it effective to study young stars\cite{Benatti2021}. In this case we obtain lower amplitudes for the signals attributed to planets \textit{b} and \textit{e}, and higher for planet \textit{c}. This was the least favoured of the models we tested (Ln$Z$ of -- 4563).  For the most favoured model (PQP2) we tested the difference between having 4 planetary components in the RV, 2 planetary components ({\it b} and {\it e}) and no planetary components. We found that a model with 2 Keplerian components in the RV is much more likely than a model with no planetary signals, with a $\Delta$ Ln$Z$ $>$ 25 (false alarm probability $<$ 0.1\%), and also more much likely than a model with only planet b , with a $\Delta$ Ln$Z$ $>$ 20 (false alarm probability $<$0.1\%). The model with 4 Keplerian components is less significant than the model with 2 Keplerian components, which is not surprising considering we could not detect the RV signals of planets {\it c} and {\it d}.

\begin{equation}
\fontsize{8}{11}\selectfont
 k(\tau) = {A^{2}} \cdot exp \Bigg[ - {\tau^2 \over{L}} - {{sin^2(\pi \tau / P_{rot})} \over{2 \omega^2} }\Bigg] 
 + (\sigma^2 (t) + \sigma^2_{j}) \cdot \delta_{\tau}
\end{equation}

\noindent As the results coming from the different GP models are slightly different, we performed simulations to test the accuracy of the amplitude measurements in this particular case. To do that we subtracted the detected planetary signals from the RV to create an "activity-only" dataset. Following the same procedure as with our original RV dataset, we tested that all the models recovered amplitudes that are consistent with zero at the periods of the planets. Later we injected planetary signals at different amplitudes to study the behaviour of every model. The results were very similar to what we had already found. The PQP2 Kernel recovered the amplitudes of the signals corresponding to  planets b  and e within a 5\% accuracy for amplitudes larger than 20 m~s$^{-1}$. The model showed a tendency to underestimate the amplitudes of planets c  and d by a 20 \% margin for amplitudes smaller than 20 m~s$^{-1}$. The combination of two SHO Kernels consistently overestimated the amplitude of planet b, while underestimating the amplitudes of the three remaining signals. The model using this Kernel also recovered smaller uncertainties for the Keplerian amplitudes in all scenarios. The QP accurately recovered the amplitudes of planets c and d, and it strongly underestimated the amplitudes of the signals corresponding to planets b and e, sometimes by a 50\% margin.  This is not a fully unexpected behaviour, as the more flexible GP Kernels have a higher rate of false negatives\cite{Feng2016}. Figure~\ref{test_Kernels} shows the comparison between the injected and recovered planetary amplitudes using the three GP Kernels. The PQP2 Kernel provided the best consistent results for all the tested combinations.

\noindent To further test our results we opted for a different approach based on the correlation of the RVs with the photometric data. In spot-dominated stars, the activity-induced RV variations are correlated with the gradient of the flux$^{83}$. This correlation can be used to detrend the data from stellar activity. As we do not have simultaneous, but contemporanous, photometry, we calculated the gradient from the model of the photometry. We modelled the rotation using a third order polynomial against the derivative of the flux. A first attempt left some residual power at the first harmonic of the rotation period, which led us to include a sinusoidal component at that period. Our activity model is then defined as: 

\begin{equation}
\fontsize{8}{11}\selectfont
RV_{rot} = C1 \cdot {{dFlux}\over{dt}} +  C2 \cdot \Bigg({{dFlux}\over{dt}}\Bigg)^2 +  C3 \cdot \Bigg({{dFlux}\over{dt}}\Bigg)^3  +   
A_{rot} \cdot sin(4 \cdot \pi {{(t - T0)}\over {P_{rot}}})
\end{equation}

\noindent where T0 was parametrised as JD$_{0}$+$P_{\rm rot}\cdot \phi$, with JD$_{0}$ = 2458791.627.

\noindent Using this model we recovered a very similar solution as with the mixture of two SHO Kernels, although with much larger residuals. We detected the presence of the planets V1298 Tau b and e, and measured upper limits for the amplitudes of planets c and d. We measure the amplitude of planet {\it e} to be much larger than what was found with the GP models, which might be caused by the Keplerian model absorbing some unmodelled activity. 

\noindent Table ED~\ref{parameters_1} shows the parameters used in the fit, the datasets involved in fitting every every parameter, the priors and the results obtained for the different models tested and Figure~\ref{corner_plot} shows the corner plot of the parameters of the most significant model. Figure~\ref{rv_ts} shows the RV with the best fit model for the raw data, activity-filtered data, planet-filtered data and residuals, along with the GLS periodogram\cite{Zechmeister2009} of each of those datasets. To represent the activity model we used a weighted average of the models for the different instruments. Figure~\ref{data_lco} shows the best fit to the contemporary photometry and Figure~\ref{data_k2} shows the best fit to the K2 observations.

\subsection{Lessons learned and limitations.}

We found that the signal phase-folded to the rotation period shows clearly the two modes of oscillation that our favoured GP Kernel describes. The amplitude of the rotation signal is 8 times larger than the amplitude measured for the signal related to V1298 Tau b, and 5 times larger than the signal related to planet e. In the context of young exoplanets, the stellar activity signals engulf those signals related to the planets and therefore similarly large observational efforts with precise RV measurements will be required.

\noindent We found that not all GP Kernels behaved the same at all timescales in our dataset. The classic QP Kernel handles short-period signals quite well. However for longer period signals it seems to absorb a significant part of the Keplerian components, causing a clear underestimation of the measured amplitudes. The mixture of SHO Kernels had the opposite behaviour. It {\it underfits} the activity component, leaving larger residuals and causing an overestimation of (some of) the Keplerian amplitudes. We found that our Kernel of choice (PQP2) provides better description of the activity variations of V1298 Tau and a more accurate determination of the Keplerian amplitudes. 

\noindent It is important to remain cautious about the mass determined for the planet V1298 Tau e. The original detection did not constrain the orbital period, which is derived purely from the RV information. We studied the S$_{MW}$ index\cite{Noyes1984}, H${\alpha}$ index\cite{GomesdaSilva2011}, NaI index\cite{Diaz2007} and TiO\cite{Azizi2018} chromospheric indicators following a very similar procedure as with the RV data, not finding any significant periodicity (aside from the rotation) at periods shorter than 150 days, which favours the planetary hypothesis.  However, disentangling planetary signals from stellar activity in RV in young stars such as V1298 Tau is a very challenging task. Without the confirmation of its orbital period by transit photometry it is very difficult to completely exclude a stellar origin (or contribution) to the signal.  

\end{methods}


\bibliographystyle{unsrt}


\begin{addendum}

\item[Code availability] The \texttt{SERVAL} template-matching radial velocity measurement tool, \texttt{Celerite}, \texttt{George}, \texttt{EMCEE}, \texttt{dynesty}, \texttt{RADVEL}, \texttt{PyTransit}, \texttt{AstroImageJ}, \texttt{SYNPLE},\texttt{StePar}, \texttt{FERRE} and \texttt{MOOG} are easily accessible open source projects. Additional software available upon request.

 \item[Data availability] The public high-resolution spectroscopic raw data used in the study can be freely downloaded from the corresponding facility archives. Proprietary raw data are available from A.S.M on reasonable request.
\end{addendum}



\begin{figure}
\centering
\includegraphics[width=0.95\textwidth,angle=0]{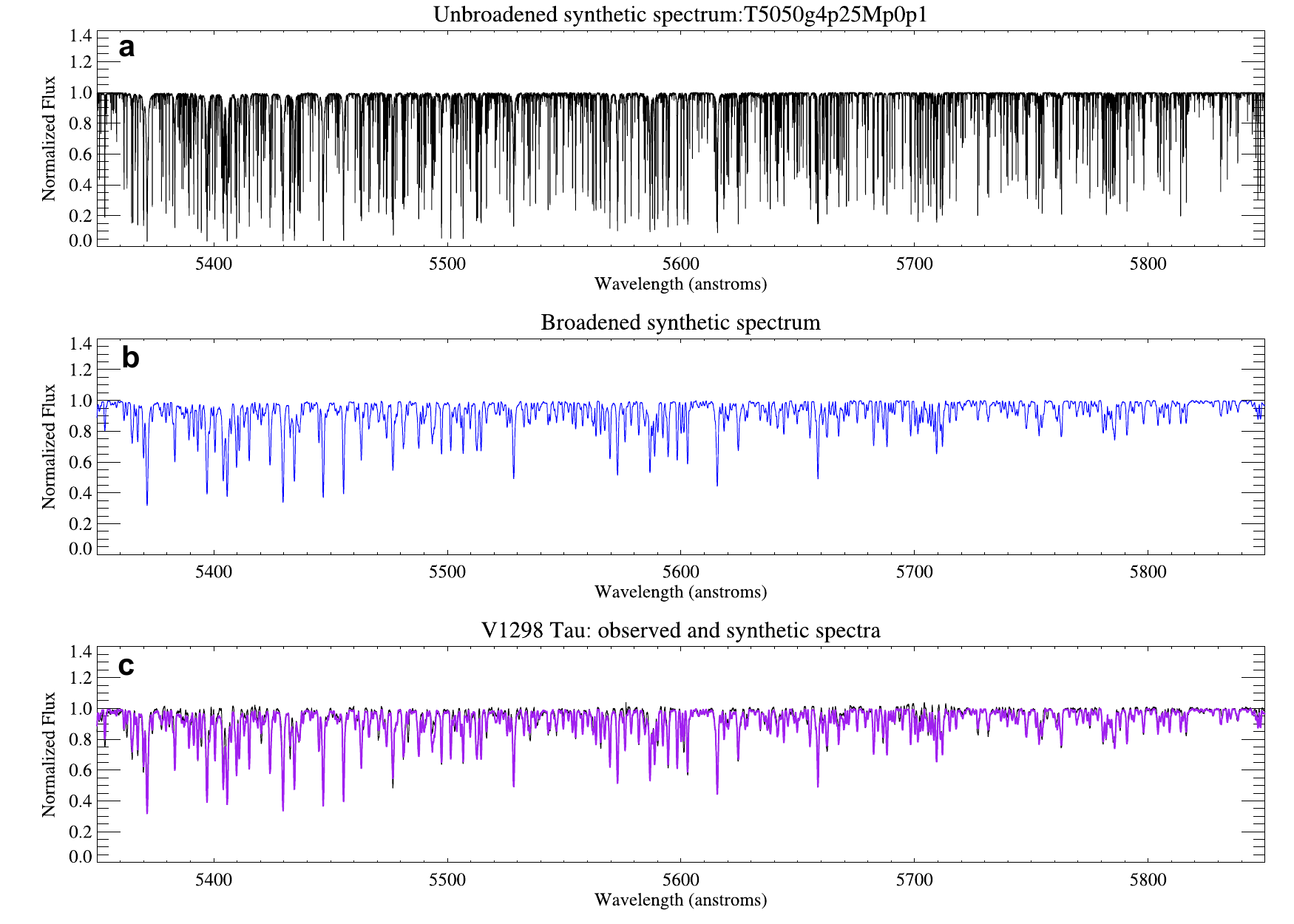}
\bigskip
\noindent
\begin{minipage}[b]{0.95\textwidth}
\caption{\textbf{Best synthetic spectral fit of the HARPS-N spectrum of V1298 Tau.} 
The interpolated SYNPLE synthetic spectrum without rotational broadening computed for the 
derived best-fit stellar parameters and metallicity (a), the broadened spectrum with a rotational velocity of 24~\kms (b) and the observed HARPS-N 1D spectrum of 
V1298 Tau (black line) together with the best-fit synthetic spectrum (purple line) 
are displayed in the spectral range 5350--5850~{\AA} (c).}\label{syn_spec}
\end{minipage}

\end{figure}


\begin{figure}
\centering
\includegraphics[width=0.95\textwidth,angle=0]{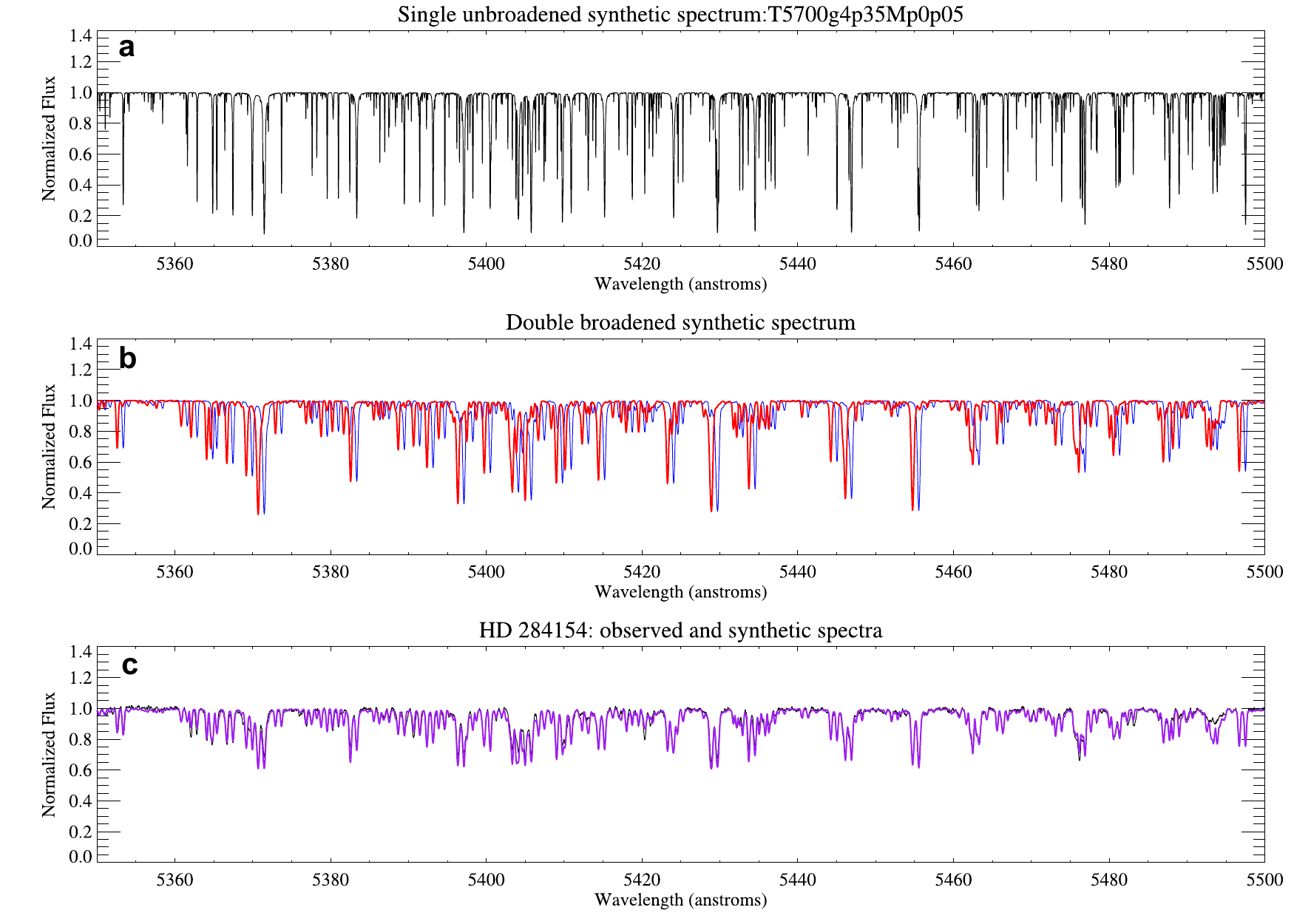}
\bigskip
\noindent
\begin{minipage}[b]{0.95\textwidth}
\caption{\textbf{Best synthetic spectral fit to the NOT FIES spectrum of HD\,284154.} 
The interpolated SYNPLE synthetic spectra without broadening computed for the derived 
best-fit stellar parameters and metallicity (a), the synthetic spectra separated 
by RV~$\sim43.6$~\kms with a rotational velocity of 10~\kms (b) and the 
observed FIES 1D spectrum of HD~284154 (black line) together with the best-fit synthetic 
double-lined spectrum (purple line) are displayed in the spectral range 5350--5850~{\AA} 
(c).} \label{spec_binary}
\end{minipage}

\label{frem}
\end{figure}

\begin{figure}
\centering
\includegraphics[width=0.95\textwidth,angle=0]{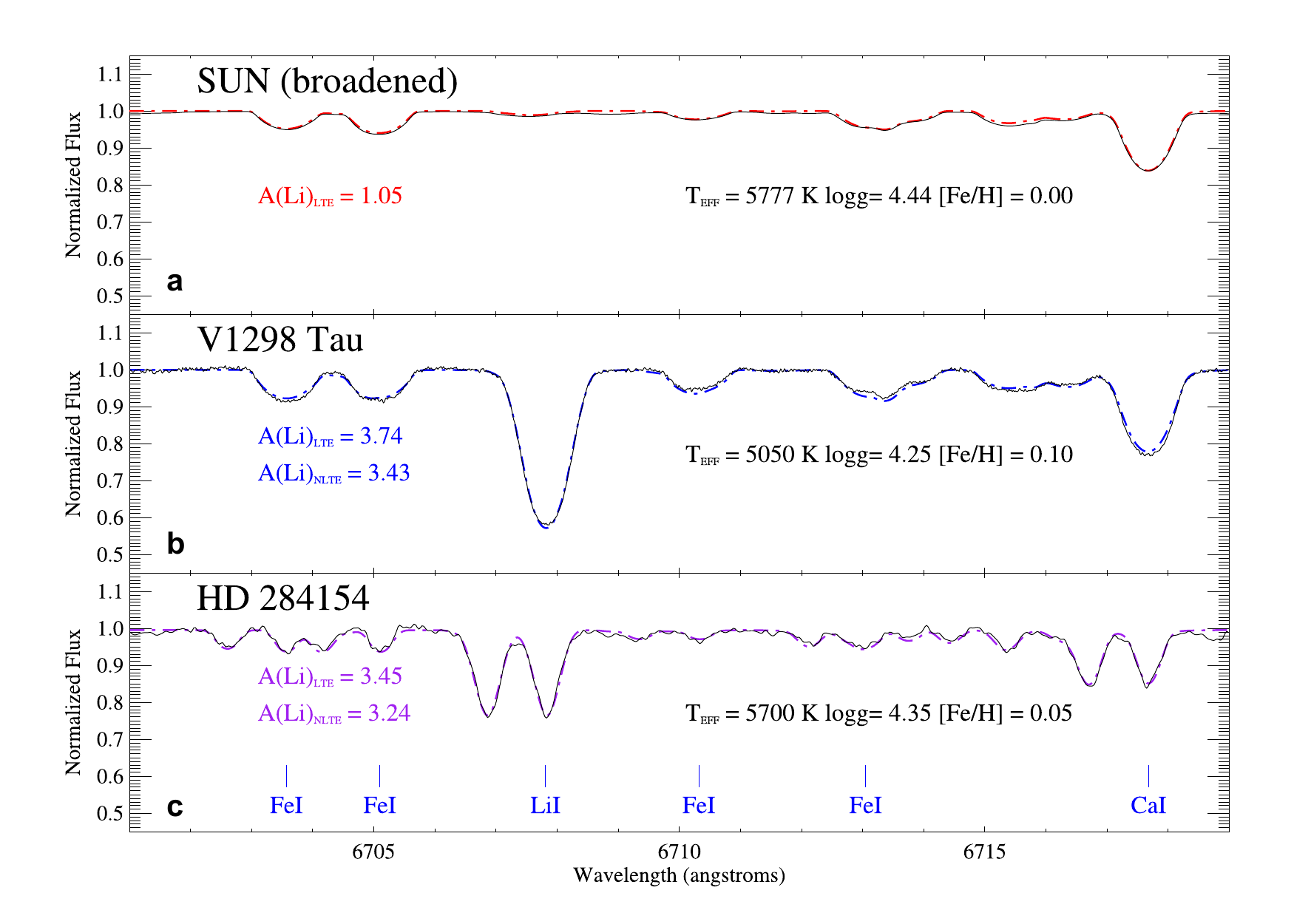}
\bigskip
\noindent
\begin{minipage}[b]{0.95\textwidth}
\caption{\textbf{The lithium spectral region of V1298 Tau and HD~284154.}
Spectral region of the lithium doublet around $6708$~{\AA} of the solar ATLAS spectrum 
broadened with a rotation profile of 24~\kms (a), the HARPS-N spectrum of 
V1298 Tau (b), and the FIES spectrum of the double-lined spectroscopic 
binary HD~284154 (c), together with the best-fit MOOG synthetic spectra.} \label{spec_li}
\end{minipage}

\end{figure}


\begin{figure}
\centering
\begin{minipage}{0.5\textwidth}
        \centering
        \includegraphics[width=8.0cm]{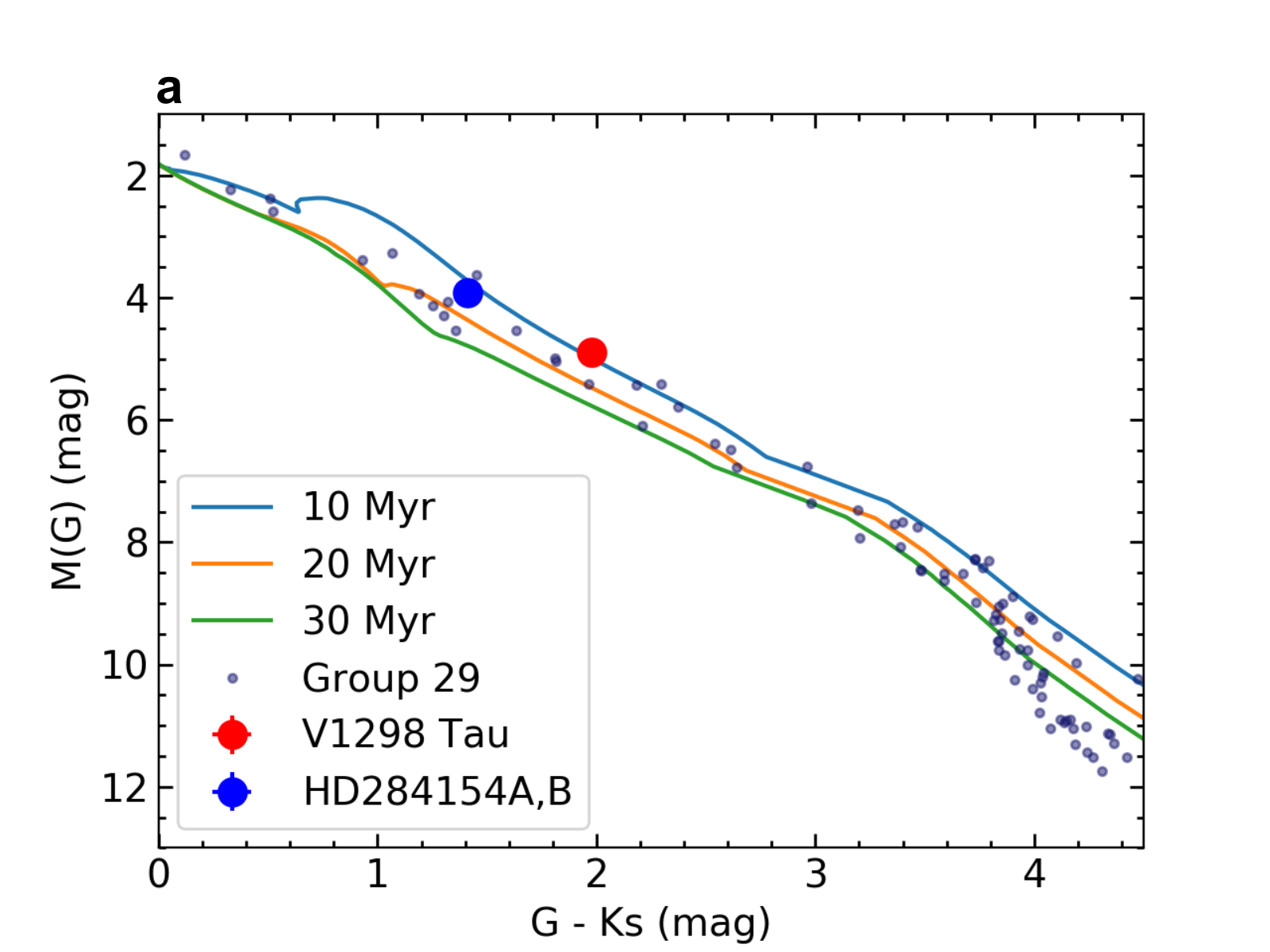}
\end{minipage}%
\begin{minipage}{0.5\textwidth}
        \centering
        \includegraphics[width=8.0cm]{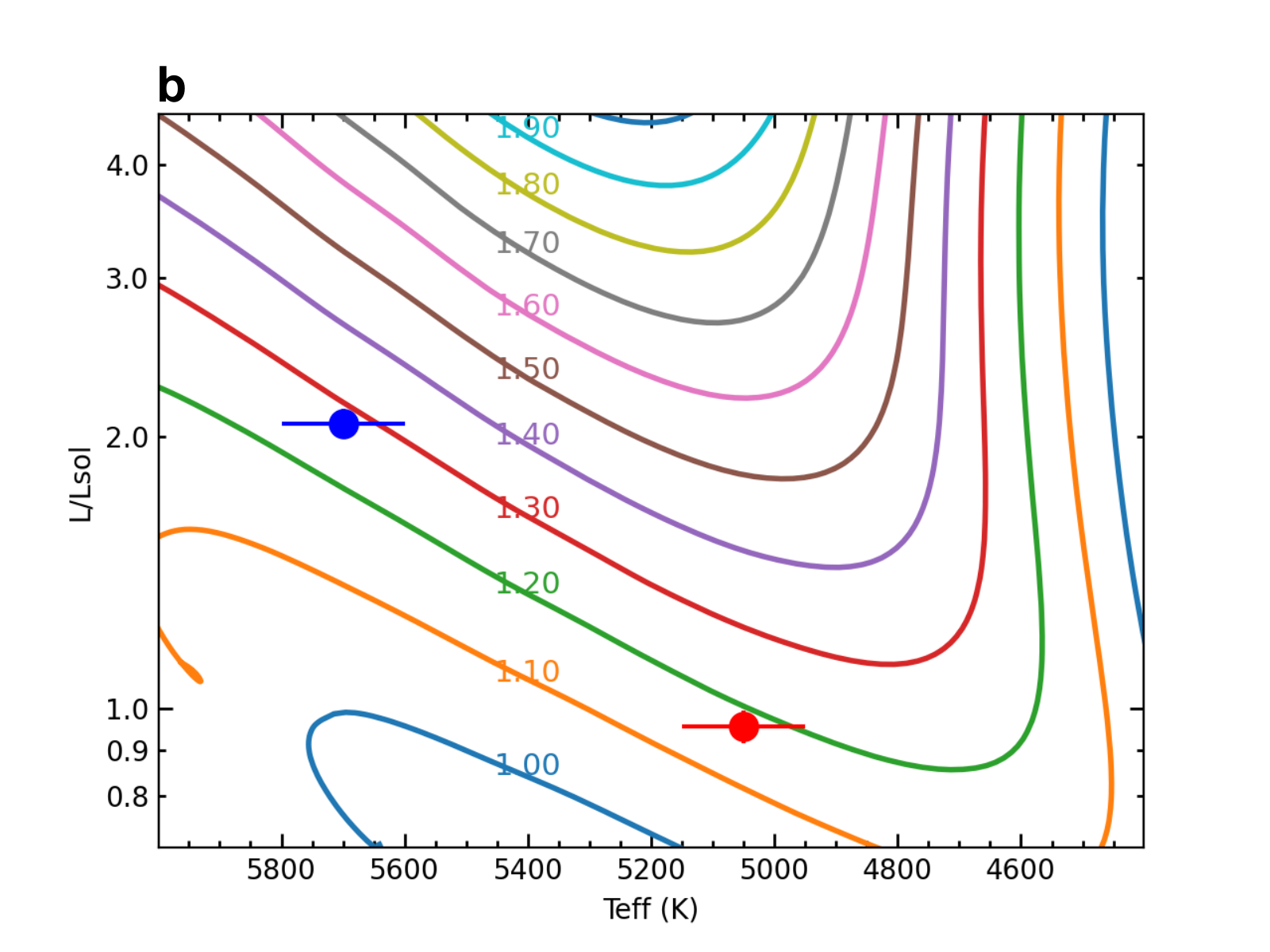}
\end{minipage}%

        \caption{\textbf{Position of V1298 Tau and HD 284154 in the  colour-magnitude and Hertzsprung-Russel diagrams.} \textit{a:} Colour-magnitude diagram of V1298 Tau and HD 284154A and B (separate components) and the other group 29 members along with various PARSEC isochrones\cite{Bressan12}. The 20-Myr isochrone nicely reproduces the sequence of stars with colours  $G-K_s < 3.5$ mag while the 10- and 30-Myr isochrones provide acceptable upper and lower envelopes to the observed dispersion of the Group 29 sequence. \textit{b:} Location of V1298 Tau (red) and HD 284154 (blue) in the Hertzsprung-Russel diagram. HD 284154 is decomposed into two equal mass and equal luminosity stars. The tracks for masses between 1.0 and 1.9 M$_\odot$ are also shown and are labeled with the mass value in solar units. Note that the luminosity axis is in logarithmic scale. The error bar in luminosity is of the size of the symbol.} 
        \label{gaia_colours}
\end{figure} 

\begin{figure}
\centering

        \includegraphics[width=\textwidth]{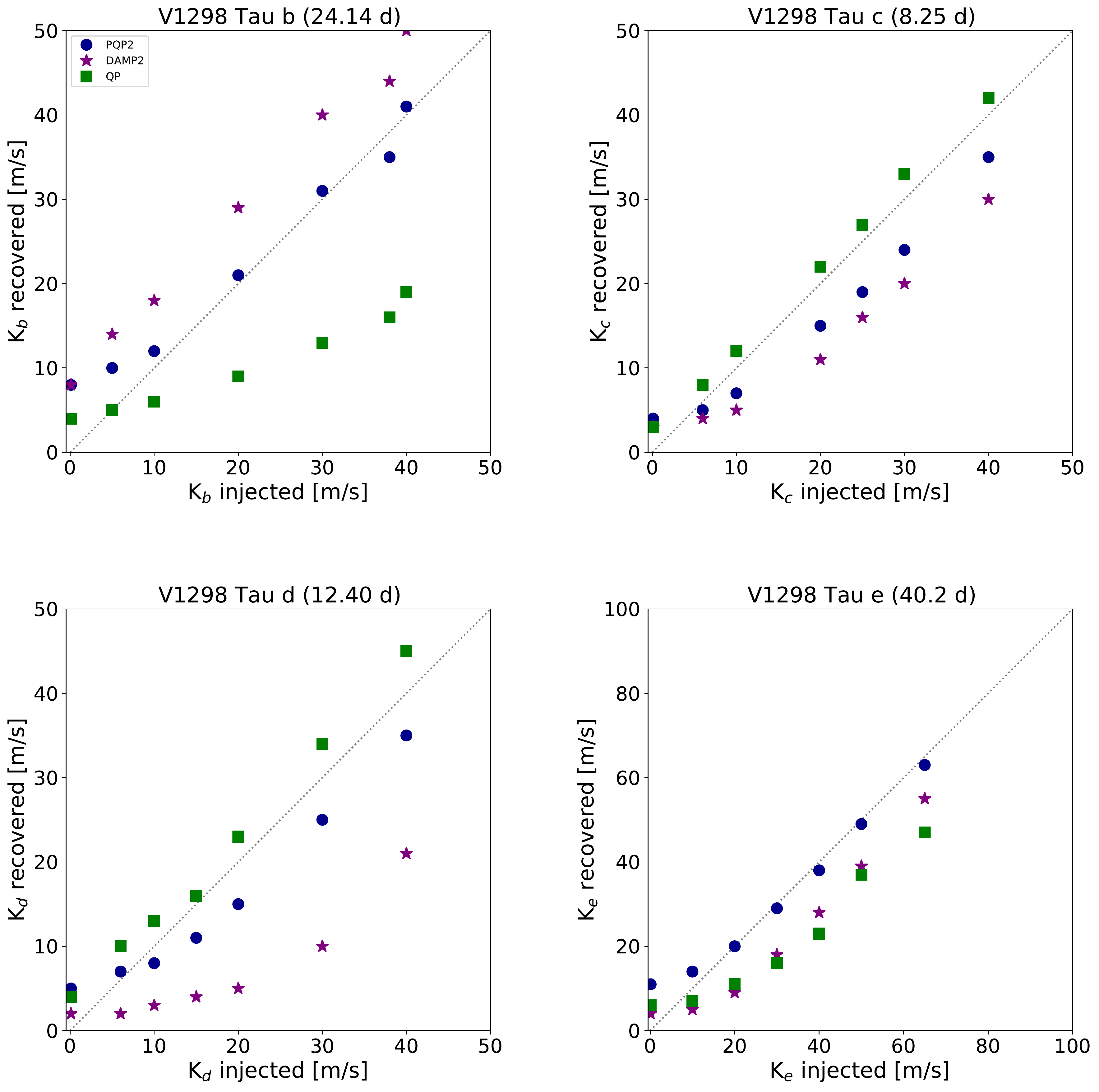}
        \caption{\textbf{Accuracy of the recovered planetary amplitudes of the different methods.} Recovered planetary amplitude against injected planetary amplitude in the simulated datasets for the four planets in the system}.
        \label{test_Kernels}
\end{figure} 

\begin{figure}
        \centering
        \includegraphics[width=\textwidth]{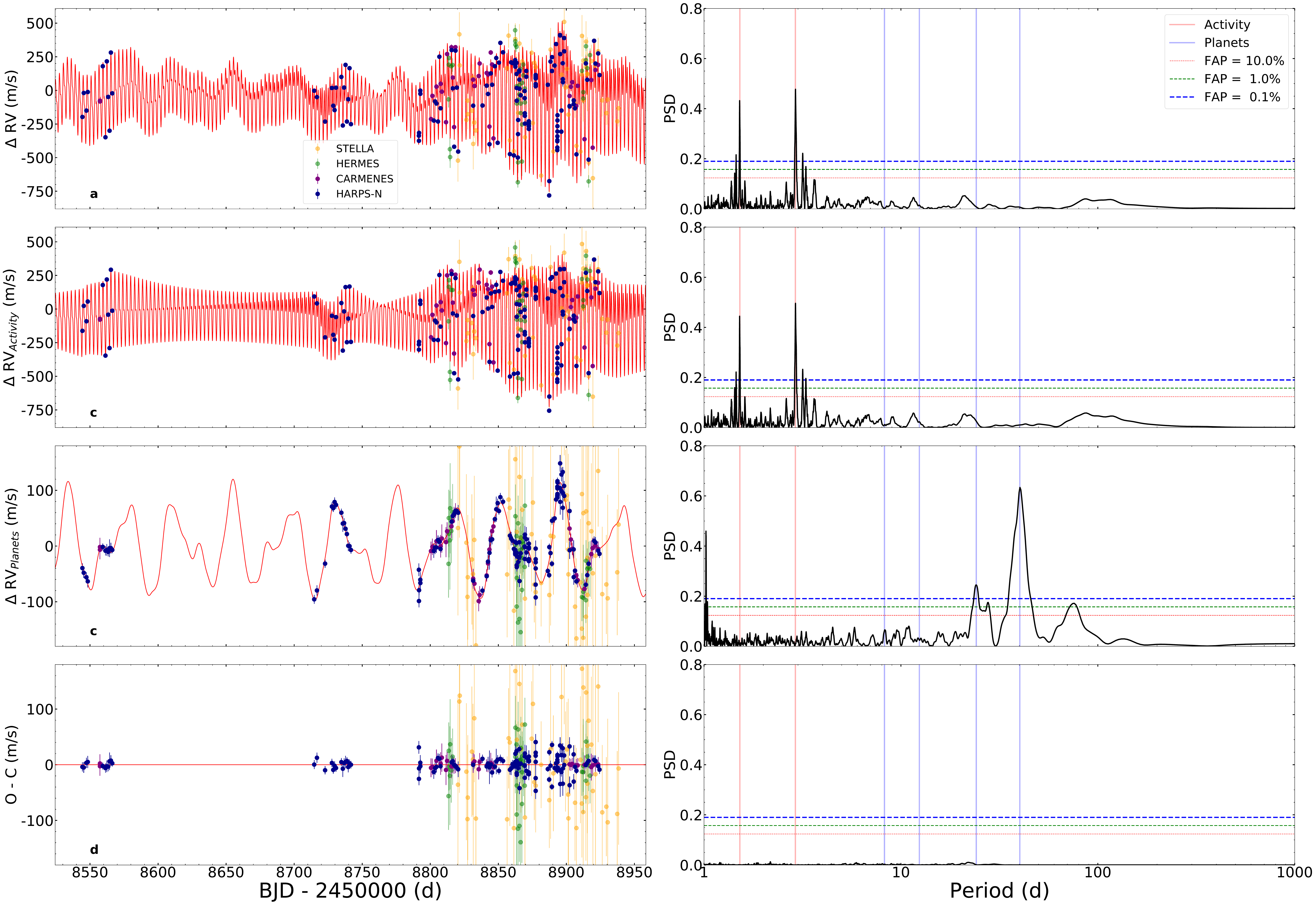}

        \caption{\textbf{RV time series with the best model fit of V1298 Tau and their periodograms.} \textit{a:} Full time series with the best fit model combining stellar activity and planetary signals. The stellar activity model represented is a weighted average of the models used for the different spectral ranges. \textit{b:} Activity induced RV after subtracting the planetary signal. \textit{c:} Planetary RV component, after subtracting the stellar induced signal. \textit{d:} Residuals after the fit. 1$\sigma$ error bars (internal RV uncertainties) of the measurements are shown. The right panel of each figure shows the periodogram of the data with their associated levels of false alarm probability. The positions of the activity, and planetary, signals are indicated with red and blue vertical lines, respectively. }
        \label{rv_ts}
\end{figure}

\begin{figure}
\centering
        \includegraphics[width=\textwidth]{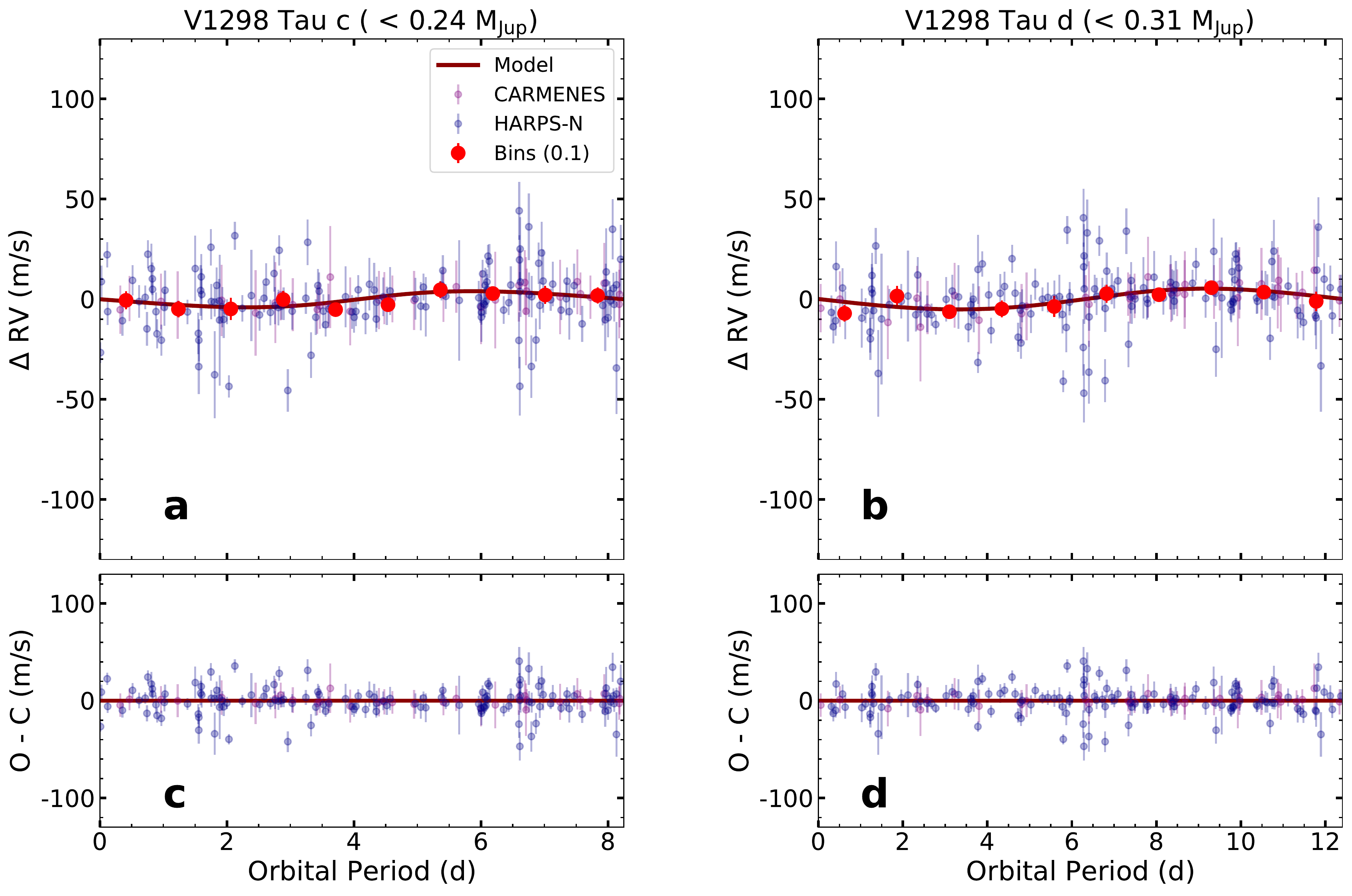}
        \caption{\textbf{Phase-folded plots of the RV signals for the two planets of the V1298\,Tau planetary system for which we could not confirm the RV signals.} \textit{a}: Phase-folded representation of the best-fitting Keplerian orbit (red line) for V1298\,Tau c. \textit{b}: Same for V1298\,Tau d. \textit{c and d}: Residuals after the fit for both cases. For a better visualisation, only HARPS and CARMENES data have been included. In all cases, 1$\sigma$ error bars (internal RV uncertainties) of the measurements are shown.}
        \label{rv_fold_2}
\end{figure} 

\begin{figure}
\centering
        \includegraphics[width=\textwidth]{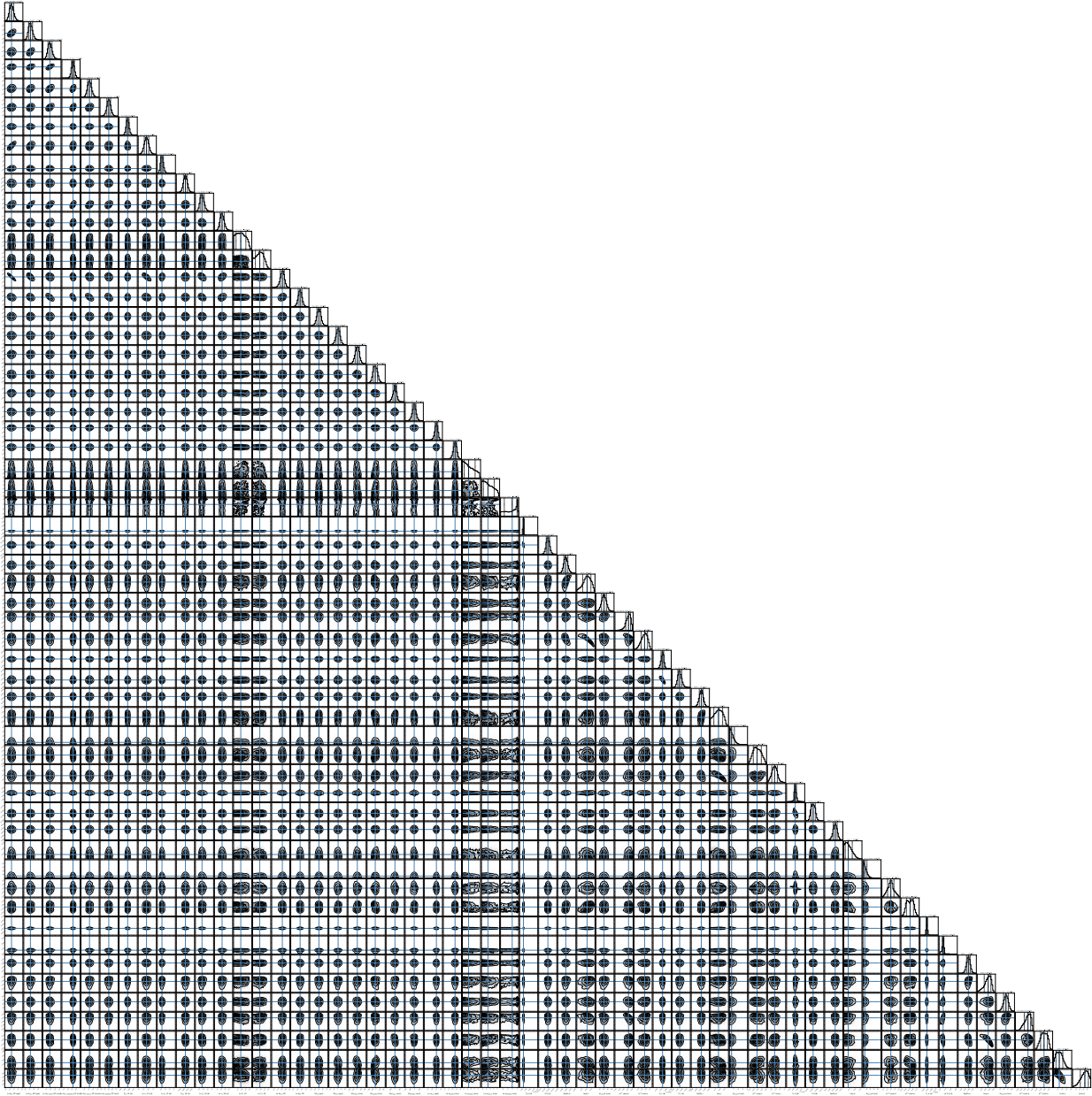}
        \caption{\textbf{Corner plot of the parameters of the best model fit (PQP2).} Posterior distributions of all the parameters sampled for the best model fit along with the correlation maps between them.}
        \label{corner_plot}
\end{figure}

\begin{figure}
\centering
        \includegraphics[width=\textwidth]{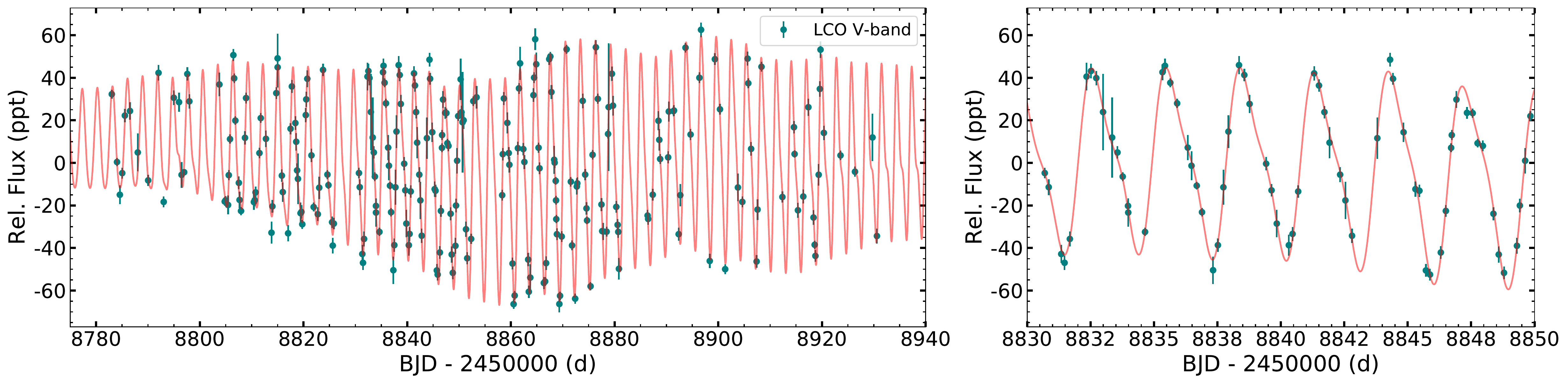}
        \caption{\textbf{LCOGT $V$-band photometry.} \textit{a:} Time series of the LCOGT $V$-band photometry with the best fit obtained from the global analysis. \textit{b:} Zoom to a well-sampled section. 1$\sigma$ error bars (internal uncertainties) of the measurements are shown.}
        \label{data_lco}
\end{figure} 

\begin{figure}
\centering
        \includegraphics[width=\textwidth]{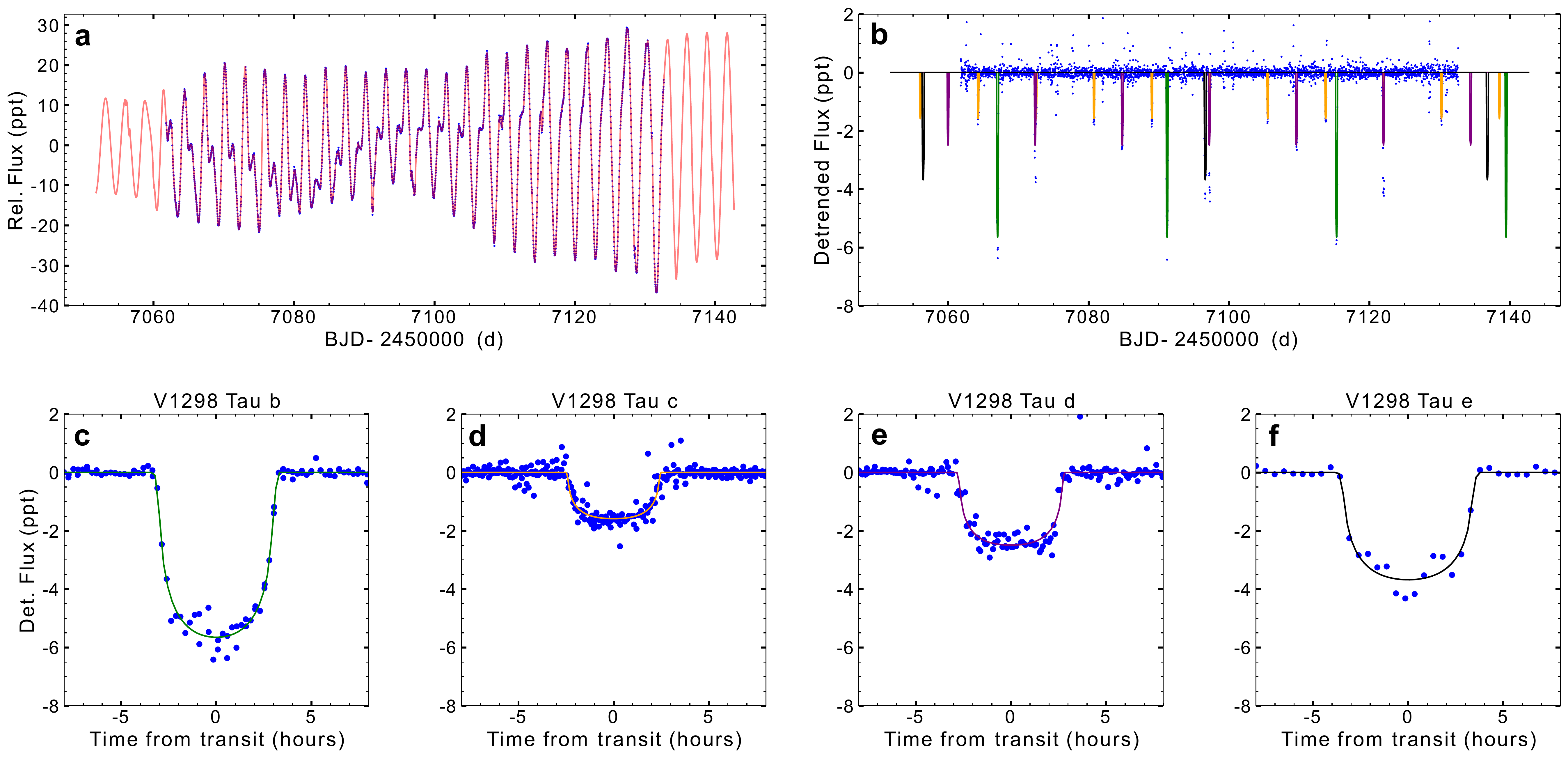}
        \caption{\textbf{K2 photometry.} Time series of the K2 photometry with the best fit obtained from the global analysis. \textit{a:} K2 data with the full fit. \textit{b:} Data detrended from stellar activity with the best fit to the transits. \textit{c,d,e and f:} Phase-folded plots of the transits of the four planets.}
        \label{data_k2}
\end{figure}

\fontsize{8}{11}\selectfont
\begin{longtable}{llccccc}
\caption{Parameters and priors for the combined model. The column "Dataset" shows between which datasets the parameter is shared during the optimisation. All T0$s$ are expressed in BJD - 2450000. Datasets: 1 - K2; 2 - LCO; 3 - HARPS-N RV; 4 -  CARMENES RV; 5 - SES RV; 6 - HERMES RV. $^{1}$The correlation model ( 4p~$_{Corr}$) uses a different amount of data. \label{parameters_1}}\\
\endfirsthead
\multicolumn{6}{c}{\tablename\ \thetable\ -- \textit{Continued from previous page}} \\
\hline
\endhead
\hline \multicolumn{7}{r}{\textit{Continued on next page}} \\
\endfoot
\hline
\endlastfoot
\hline

\textbf{Parameter} & \textbf{Dataset} & \textbf{Priors} & \textbf{4p~$_{PQP2}$} & \textbf{4p~$_{Damped}$} & \textbf{4p~$_{QP}$} & \textbf{4p~$_{Corr}$}\\ \hline

\textit{Planets} \\ 
T0~b [d] & 1,3,4,5,6 & $\mathcal{N}$ (7067.0488, 0.25) & 7067.0486$^{+0.0015}_{-0.0016}$ &7067.0485$^{+0.0016}_{-0.0017}$ &7067.0488$^{+0.0016}_{-0.0017}$ &7067.0484$^{+0.0032}_{-0.0038}$ \\ 
P~b  [d] & 1,3,4,5,6 & $\mathcal{N}$ (24.1396, 0.25) & 24.1399$^{+0.0016}_{-0.0015}$ &24.14$^{+0.0016}_{-0.0016}$ &24.1396$^{+0.0016}_{-0.0016}$ &24.14$^{+0.0034}_{-0.0031}$ \\ 
R$_{p}$/R$_{*}$~b  & 1 &  $\mathcal{U}$ (0 , 0.2) & 0.0698$^{+0.0024}_{-0.0023}$ &0.0711$^{+0.0029}_{-0.0029}$ &0.0715$^{+0.0024}_{-0.0024}$ &0.0695$^{+0.0037}_{-0.0038}$  \\ 
Imp~b & 1 &  $\mathcal{U}$ (0 , 1) & 0.33$^{+0.12}_{-0.14}$ &0.45$^{+0.08}_{-0.16}$ &0.45$^{+0.06}_{-0.08}$ &0.32$^{+0.1}_{-0.15}$ \\ 
K~b [m$\cdot$s$^{-1}$] & 3,4,5,6&  $\mathcal{U}$ (0 , 200) & 41$^{+12}_{-12}$ &54.6$^{+7.6}_{-7.8}$ &31$^{+18}_{-17}$ &57$^{+20}_{-20}$ \\ 
$\sqrt{e}$ $\cdot$ cos $\omega$~b & 1,3,4,5,6&  $\mathcal{U}$ (0 , 0.5) & 0.31$^{+0.12}_{-0.16}$ &0.35$^{+0.07}_{-0.09}$ &0.02$^{+0.22}_{-0.22}$ &0.05$^{+0.23}_{-0.22}$ \\ 
$\sqrt{e}$ $\cdot$ sin $\omega$~b & 1,3,4,5,6&  $\mathcal{U}$ (0 , 0.5) & -0.06$^{+0.14}_{-0.17}$ &-0.09$^{+0.19}_{-0.17}$ &-0.0$^{+0.17}_{-0.2}$ &-0.01$^{+0.16}_{-0.18}$ \\ \\ 
T0~c [d] & 1,3,4,5,6 & $\mathcal{N}$ (7064.2797, 0.25) & 7064.2801$^{+0.0039}_{-0.0046}$ &7064.2806$^{+0.0037}_{-0.0042}$ &7064.2785$^{+0.0045}_{-0.005}$ &7064.2778$^{+0.0081}_{-0.0084}$ \\ 
P~c  [d] & 1,3,4,5,6 & $\mathcal{N}$ (8.24958, 0.25) & 8.2492$^{+0.001}_{-0.0008}$ &8.249$^{+0.00094}_{-0.00081}$ &8.2508$^{+0.0014}_{-0.0013}$ &8.2498$^{+0.0018}_{-0.0016}$ \\ 
R$_{p}$/R$_{*}$~c  & 1 &  $\mathcal{U}$ (0 , 0.2) & 0.0371$^{+0.0019}_{-0.0019}$ &0.0372$^{+0.0017}_{-0.0017}$ &0.0368$^{+0.002}_{-0.002}$ &0.0361$^{+0.0031}_{-0.0035}$ \\ 
Imp~c & 1 &  $\mathcal{U}$ (0 , 1) & 0.26$^{+0.13}_{-0.15}$ &0.25$^{+0.14}_{-0.15}$ &0.22$^{+0.15}_{-0.14}$ &0.17$^{+0.16}_{-0.11}$ \\ 
K~c [m$\cdot$s$^{-1}$] & 3,4,5,6&  $\mathcal{U}$ (0 , 200) & 4.0$^{+4.9}_{-2.9}$ &3.6$^{+4.3}_{-2.6}$ &10.1$^{+4.0}_{-4.2}$ &14.3$^{+12.9}_{-9.7}$ \\ 
$\sqrt{e}$ $\cdot$ cos $\omega$~c & 1,3,4,5,6&  $\mathcal{U}$ (0 , 0.5) & -0.03$^{+0.26}_{-0.22}$ &0.01$^{+0.25}_{-0.24}$ &0.32$^{+0.12}_{-0.2}$ &-0.03$^{+0.18}_{-0.2}$ \\ 
$\sqrt{e}$ $\cdot$ sin $\omega$~c & 1,3,4,5,6&  $\mathcal{U}$ (0 , 0.5) & -0.24$^{+0.12}_{-0.12}$ &-0.11$^{+0.16}_{-0.16}$ &-0.15$^{+0.11}_{-0.12}$ &-0.15$^{+0.18}_{-0.13}$ \\ \\ 
T0~d [d] & 1,3,4,5,6 & $\mathcal{N}$ (7072.3913, 0.25) & 7072.3907$^{+0.0063}_{-0.0045}$ &7072.3902$^{+0.0038}_{-0.0033}$ &7072.3956$^{+0.0053}_{-0.0058}$ &7072.3903$^{+0.0071}_{-0.0061}$ \\ 
P~d  [d] & 1,3,4,5,6 & $\mathcal{N}$ (12.4032, 0.25) & 12.4054$^{+0.0018}_{-0.0017}$ &12.4053$^{+0.0019}_{-0.0017}$ &12.4047$^{+0.003}_{-0.0019}$ &12.4058$^{+0.0027}_{-0.0026}$ \\ 
R$_{p}$/R$_{*}$~d  & 1 &  $\mathcal{U}$ (0 , 0.2) & 0.0462$^{+0.0021}_{-0.0021}$ &0.0464$^{+0.002}_{-0.0021}$ &0.0459$^{+0.0024}_{-0.0025}$ &0.046$^{+0.0037}_{-0.004}$ \\ 
Imp~d & 1 &  $\mathcal{U}$ (0 , 1) & 0.12$^{+0.11}_{-0.08}$ &0.18$^{+0.14}_{-0.12}$ &0.21$^{+0.15}_{-0.14}$ &0.19$^{+0.13}_{-0.12}$ \\ 
K~d [m$\cdot$s$^{-1}$] & 3,4,5,6&  $\mathcal{U}$ (0 , 200) & 5.2$^{+5.9}_{-3.7}$ &2.2$^{+3.0}_{-1.6}$ &4.6$^{+4.3}_{-3.1}$ &7.0$^{+9.2}_{-5.0}$ \\ 
$\sqrt{e}$ $\cdot$ cos $\omega$~d & 1,3,4,5,6&  $\mathcal{N}$ (0 , 0.5) & -0.01$^{+0.16}_{-0.16}$ &-0.03$^{+0.23}_{-0.21}$ &0.03$^{+0.17}_{-0.2}$ &-0.04$^{+0.2}_{-0.19}$ \\ 
$\sqrt{e}$ $\cdot$ sin $\omega$~d & 1,3,4,5,6&  $\mathcal{N}$ (0 , 0.5) & -0.1$^{+0.13}_{-0.14}$ &-0.06$^{+0.16}_{-0.13}$ &0.04$^{+0.14}_{-0.15}$ &-0.17$^{+0.15}_{-0.11}$ \\ \\ 
T0~e [d] & 1,3,4,5,6 & $\mathcal{N}$ (7096.6229, 0.25) & 7096.6227$^{+0.0033}_{-0.0032}$ &7096.6227$^{+0.0032}_{-0.003}$ &7097.1913$^{+0.0018}_{-0.0036}$ &7096.6234$^{+0.0089}_{-0.0083}$ \\ 
Ln P~e  [d] & 1,3,4,5,6 & $\mathcal{U}$ (Ln(35),Ln(400)) & 3.693$^{+0.023}_{-0.023}$ &3.65$^{+0.043}_{-0.022}$ &3.718$^{+0.049}_{-0.026}$ &3.717$^{+0.048}_{-0.046}$ \\ 
R$_{p}$/R$_{*}$~e  & 1 &  $\mathcal{U}$ (0 , 0.2) & 0.0592$^{+0.0046}_{-0.0048}$ &0.0599$^{+0.0045}_{-0.0047}$ &0.0695$^{+0.0045}_{-0.0048}$ &0.0542$^{+0.0096}_{-0.0131}$ \\ 
Imp~e & 1 &  $\mathcal{U}$ (0 , 1) & 0.41$^{+0.1}_{-0.15}$ &0.45$^{+0.08}_{-0.15}$ &0.52$^{+0.09}_{-0.13}$ &0.4$^{+0.16}_{-0.21}$ \\ 
K~e [m$\cdot$s$^{-1}$] & 3,4,5,6&  $\mathcal{U}$ (0 , 200) & 62$^{+15}_{-16}$ &59.4$^{+7.8}_{-7.7}$ &36$^{+13}_{-14}$ &80$^{+20}_{-20}$ \\ 
$\sqrt{e}$ $\cdot$ cos $\omega$~e & 1,3,4,5,6&  $\mathcal{U}$ (0 , 0.5) & 0.22$^{+0.18}_{-0.25}$ &0.21$^{+0.16}_{-0.23}$ &-0.02$^{+0.32}_{-0.26}$ &0.16$^{+0.2}_{-0.26}$ \\ 
$\sqrt{e}$ $\cdot$ sin $\omega$~e & 1,3,4,5,6&  $\mathcal{U}$ (0 , 0.5) & -0.03$^{+0.2}_{-0.2}$ &0.04$^{+0.2}_{-0.22}$ &0.27$^{+0.13}_{-0.26}$ &0.04$^{+0.19}_{-0.24}$ \\ \\ 
\textit{Activity} \\ 
Ln A [ppt] & 1 & $\mathcal{U}$ (-10 , 10) & 2.71$^{+0.27}_{-0.25}$ &2.85$^{+0.33}_{-0.3}$ &2.95$^{+0.64}_{-0.46}$ &3.16$^{+0.93}_{-0.68}$ \\ 
Ln A [ppt] & 2 & $\mathcal{U}$ (-10 , 10) & 3.27$^{+0.2}_{-0.17}$ &3.32$^{+0.23}_{-0.28}$ &3.45$^{+0.23}_{-0.2}$ &\\ 
Ln A RV [m$\cdot$s$^{-1}$] & 3 & $\mathcal{U}$ (-10 , 10) & 5.53$^{+0.19}_{-0.17}$ &5.28$^{+0.19}_{-0.18}$ &5.44$^{+0.12}_{-0.11}$ &-3.25$^{+4.21}_{-4.49}$ \\ 
Ln A RV [m$\cdot$s$^{-1}$] & 4 & $\mathcal{U}$ (-10 , 10) & 5.54$^{+0.22}_{-0.2}$ &5.33$^{+0.26}_{-0.39}$ &5.46$^{+0.16}_{-0.15}$ &-3.1$^{+4.15}_{-3.94}$ \\ 
Ln A RV [m$\cdot$s$^{-1}$] & 5 & $\mathcal{U}$ (-10 , 10) & 5.58$^{+0.23}_{-0.22}$ &5.21$^{+0.24}_{-0.22}$ &5.65$^{+0.16}_{-0.16}$ &-3.05$^{+4.74}_{-4.23}$ \\ 
Ln A RV [m$\cdot$s$^{-1}$] & 6 & $\mathcal{U}$ (-10 , 10) & 6.01$^{+0.25}_{-0.26}$ &5.68$^{+0.26}_{-0.23}$ &5.98$^{+0.19}_{-0.17}$ &-3.96$^{+4.86}_{-4.09}$ \\ 
P$_{rot}$ 2015 [d] & 1 & $\mathcal{U}$ (2.75 , 3.1) & 2.868$^{+0.013}_{-0.013}$ &2.868$^{+0.013}_{-0.012}$ &2.867$^{+0.013}_{-0.012}$ &2.871$^{+0.027}_{-0.026}$ \\ 
Ln T$_{Scale~1 }$ 2015 [d] & 1 & $\mathcal{U}$ (-10 , 10) & 5.41$^{+0.69}_{-0.67}$ &5.7$^{+0.76}_{-0.76}$ &6.04$^{+1.33}_{-1.03}$ &6.09$^{+1.88}_{-1.93}$ \\ 
Ln T$_{Scale~2 }$ 2015 [d] & 1 & $\mathcal{U}$ (-10 , 10) & 0.75$^{+0.21}_{-0.18}$ &0.73$^{+0.19}_{-0.16}$ &0.64$^{+0.2}_{-0.17}$ &0.83$^{+0.58}_{-0.33}$ \\ 
P$_{rot}$ 2019 [d] & 2,3,4,5,6 & $\mathcal{U}$ (2.75 , 3.1) & 2.9101$^{+0.002}_{-0.002}$ &2.9103$^{+0.0019}_{-0.0017}$ &2.9001$^{+0.0028}_{-0.0028}$ &2.887$^{+0.1534}_{-0.1646}$ \\ 
Ln T$_{Scale~1 }$ 2019 [d] & 2,3,4,5,6 & $\mathcal{U}$ (-10 , 10) & 5.34$^{+0.4}_{-0.35}$ &4.05$^{+0.85}_{-0.47}$ &3.26$^{+0.08}_{-0.09}$ &\\ 
Ln T$_{Scale~2 }$ 2019 [d] & 2,3,4,5,6 & $\mathcal{U}$ (-10 , 10) & 4.98$^{+0.45}_{-0.4}$ &5.91$^{+0.64}_{-0.59}$ & & \\ \\ 
Ln Mix RV  & 3,4,5,6 & $\mathcal{U}$ (-10 , 10) & -0.19$^{+0.21}_{-0.21}$ &-0.12$^{+0.25}_{-0.23}$ & & \\ 
Ln Mix Phot  & 1,2 & $\mathcal{U}$ (-10 , 10) & -0.91$^{+0.24}_{-0.27}$ &-1.06$^{+0.3}_{-0.33}$ &-1.15$^{+0.46}_{-0.64}$ &-1.35$^{+0.79}_{-0.9}$ \\ 
Ln C RV [d] & 3,4,5,6 & $\mathcal{U}$ (-10 , 10) & -5.5$^{+3.2}_{-3.0}$ & & & \\ 
Ln C Phot [d] & 2 & $\mathcal{U}$ (-10 , 10) & -5.5$^{+2.5}_{-2.8}$ & & & \\ 
Ln $\omega$~RV & 3,4,5,6 & $\mathcal{U}$ (-10 , 10) &  & &-1.21$^{+0.08}_{-0.08}$ &\\ 
Ln $\omega$~Phot & 2 & $\mathcal{U}$ (-10 , 10) &  & &-0.34$^{+0.17}_{-0.16}$ &\\ 
Phase Rot & 3,4,5,6 & $\mathcal{U}$ (0 , 1) &  & & &0.58$^{+0.25}_{-0.31}$ \\ \\ 
C1 [m$\cdot$ppm$^{-1}$] & & $\mathcal{N}$(0,100)  &  & && -153.29$^{+22.93}_{-23.07}$ \\ 
C2 [m$^{2}\cdot$ppm$^{-2}$] & & $\mathcal{N}$(0,100)  &  & && 8.68$^{+13.36}_{-13.57}$ \\ 
C3 [m$^{3}\cdot$ppm$^{-3}$] & & $\mathcal{N}$(0,100)  &  & && -9.8$^{+7.48}_{-7.63}$ \\ \\ 
\textit{Instrumental} \\ 
F0$_{K2}$ [ppt] & 1 & $\mathcal{N}$ (0 , 30) & 0.01$^{+0.25}_{-0.25}$ &0.02$^{+0.24}_{-0.24}$ &0.01$^{+0.26}_{-0.26}$ &-0.01$^{+0.51}_{-0.5}$ \\ 
F0$_{LCO}$ [ppt] & 1 & $\mathcal{N}$ (0 , 30) & 2.4$^{+7.4}_{-8.1}$ &1.74$^{+0.59}_{-0.56}$ &4$^{+14}_{-13}$ &\\ 
V0$_{HARPS-N}$ RV [m$\cdot$s$^{-1}$]& 3 & $\mathcal{N}$ (0 , 100) &12$^{+77}_{-82}$ &-40.1$^{+5.9}_{-6.3}$ &-53$^{+53}_{-53}$ &-54$^{+25}_{-24}$ \\ 
V0$_{CARMENES}$ RV [m$\cdot$s$^{-1}$]& 4 & $\mathcal{N}$ (0 , 100) &-34$^{+92}_{-83}$ &5$^{+12}_{-13}$ &-44$^{+73}_{-74}$ &-6$^{+35}_{-37}$ \\ 
V0$_{SES}$ RV [m$\cdot$s$^{-1}$]& 5 & $\mathcal{N}$ (0 , 100) &-2$^{+68}_{-84}$ &0$^{+18}_{-17}$ &-52$^{+94}_{-94}$ &-11$^{+39}_{-34}$ \\ 
V0$_{HERMES}$ RV [m$\cdot$s$^{-1}$]& 6 & $\mathcal{N}$ (0 , 100) &-16$^{+75}_{-79}$ &-49$^{+16}_{-17}$ &-85$^{+136}_{-137}$ &-39$^{+38}_{-44}$ \\ \\ 
Ln Jit$_{LCO}$ Flux [ppt] & 2 & $\mathcal{U}$ (-10 , 10) & 1.2$^{+0.13}_{-0.15}$ &1.54$^{+0.35}_{-0.27}$ &1.579$^{+0.075}_{-0.075}$ &\\ 
Ln Jit$_{HARPS-N}$ RV [m$\cdot$s$^{-1}$] & 3 &  $\mathcal{U}$ (-10 , 10) & 3.05$^{+0.17}_{-0.18}$ &3.56$^{+0.22}_{-0.29}$ &2.75$^{+0.17}_{-0.19}$ &4.45$^{+0.14}_{-0.14}$ \\ 
Ln Jit$_{CARMENES}$ RV [m$\cdot$s$^{-1}$] & 4 &  $\mathcal{U}$ (-10 , 10) &-4.9$^{+5.4}_{-3.7}$ &2.6$^{+1.8}_{-7.2}$ &-5.3$^{+4.4}_{-3.1}$ &4.8$^{+0.3}_{-0.2}$ \\ 
Ln Jit$_{SES}$ RV [m$\cdot$s$^{-1}$] & 5 &  $\mathcal{U}$ (-10 , 10) &-4.7$^{+5.1}_{-3.7}$ &-1.8$^{+4.6}_{-5.3}$ &-3.5$^{+4.6}_{-4.1}$ &4.3$^{+0.6}_{-2.8}$ \\ 
Ln Jit$_{HERMES}$ RV [m$\cdot$s$^{-1}$] & 6 &  $\mathcal{U}$ (-10 , 10) & -0.3$^{+4.1}_{-6.6}$ &0.5$^{+3.3}_{-6.2}$ &-5.4$^{+5.2}_{-3.2}$ &4.8$^{+0.3}_{-0.3}$ \\ \\ 
\textit{Limg Darkening} \\ 
Limb$_{L}$  & 1 &  $\mathcal{U}$ (0 , 1) & 0.26$^{+0.16}_{-0.15}$ &0.36$^{+0.16}_{-0.15}$ &0.42$^{+0.12}_{-0.14}$ &0.44$^{+0.17}_{-0.16}$ \\ 
Limb$_{Q}$  & 1 &  $\mathcal{U}$ (0 , 1) & 0.67$^{+0.19}_{-0.25}$ &0.58$^{+0.26}_{-0.33}$ &0.23$^{+0.24}_{-0.16}$ &0.36$^{+0.26}_{-0.21}$ \\ \\ 
\textit{Residuals} \\ 

RMS O - C [m~s$^{-1}$] & 3,4,5,6 &  & 45 & 66 & 48 & 98\\
RMS O - C [m~s$^{-1}$] & 3,4 & & 26  & 14 & 18 & 65\\ \\
\textit{Bayesian Evidence} \\ 
Ln$Z$ & & & -- 4472 & -- 4549 & -- 4563 & --3488$^{1}$\\

 \hline
\end{longtable}

\end{document}